\documentclass[12pt]{article}
\usepackage{dcolumn}
\usepackage{bm}
\usepackage{a4wide}
\usepackage{fullpage}
\usepackage{slashbox}
\usepackage{tabularx} 
\usepackage{epsf}
\usepackage{amsmath,amsfonts,amssymb}
\usepackage[dvips]{graphicx}
\usepackage{array}
\newcounter{definition}

\def\thefigure{\arabic{section}.\arabic{figure}}
\def\theequation{\thesection.\arabic{equation}}
\def\appendix{
  \setcounter{section}{0}
  \setcounter{subsection}{0}
  \par
  \def\thesection{Appendix \Alph{section}}
  \def\theequation{\Alph{section}.\arabic{equation}}
  \def\thefigure{\Alph{section}.\arabic{figure}}
}
\newcommand{\nc}{\newcommand}
\nc{\rnc}{\renewcommand}
\rnc{\arraystretch}{1.3}
\nc{\xxx}{$XXX \,$}
\nc{\xxz}{$XXZ \,$}
\nc{\nn}{\nonumber}
\nc{\db}{\displaybreak[0]\\}
\nc{\ch}{\cosh}
\nc{\sh}{\sinh}
\rnc{\th}{\tanh}
\nc{\lam}{\lambda}
\nc{\kp}{\kappa}
\nc{\bra}{\langle}
\nc{\ket}{\rangle}
\rnc{\(}{\left(}
\rnc{\)}{\right)}
\nc{\lams}[2]{\lam_{#1},\ldots,\lam_{#2}}
\nc{\is}[2]{i_{#1},\ldots,i_{#2}}
\nc{\Mlams}[2]{-\lam_{#1},\ldots,-\lam_{#2}}
\nc{\lamsB}[1]
{\lam_1,\lam_2|\ldots|\lam_{2l-1},\lam_{2l}||\lam_{2l+1},\ldots,\lam_{#1}}
\nc{\isB}[1]
{i_1,i_2|\ldots|i_{2l-1},i_{2l}||i_{2l+1},\ldots,i_{#1}}
\nc{\ft}[1]{\(\frac{1+\kp}{2}\)^{#1}}
\nc{\al}[1]{\alpha(\lam_{#1})}
\nc{\g}[1]{\gamma_{\mp}(\lam_{#1})}
\nc{\hl}[1]{\lam_{#1} \rightarrow 0}
\nc{\hls}[2]{\hl{#1},\ldots,\hl{#2}}
\nc{\il}[1]{\lam_{#1} \rightarrow i \infty}
\nc{\ep}[1]{\epsilon_{#1}}
\nc{\epD}[1]{\epsilon'_{#1}}
\nc{\Tep}[1]{\tilde{\epsilon}_{#1}}
\nc{\TepD}[1]{\tilde{\epsilon}'_{#1}}
\nc{\eps}[2]{\ep{#1},\ldots,\ep{#2}}
\nc{\epDs}[2]{\epD{#1},\ldots,\epD{#2}}
\nc{\Meps}[2]{-\ep{#1},\ldots,-\ep{#2}}
\nc{\MepDs}[2]{-\epD{#1},\ldots,-\epD{#2}}
\nc{\Pk}[1]{P^{\kp}_{#1}}
\nc{\Ak}[1]{A^{\kp}_{#1}}
\nc{\Qk}[1]{Q^{\kp}_{#1}}
\nc{\ck}[1]{c^{\kp}_{#1}}
\nc{\zcor}[2]{\bra S_{#1}^z S_{#2}^z \ket}
\nc{\zzcor}[1]{2^#1 \, \bra \prod_{j=1}^#1 S_j^z \ket}
\nc{\num}[1]{\times 10^{#1}}
\title{
Correlation functions of the spin-1/2 
anti-ferromagnetic Heisenberg chain: 
exact calculation via the generating function 
}
\author{
  Jun Sato           ${}^1$ \thanks{junji@issp.u-tokyo.ac.jp},   \ \
  Masahiro Shiroishi ${}^1$ \thanks{siroisi@issp.u-tokyo.ac.jp}, \ \ 
  Minoru Takahashi   ${}^1$ \thanks{mtaka@issp.u-tokyo.ac.jp}
\\
\\ 
\it ${}^1$
Institute for Solid State Physics, University of Tokyo,\\\it 
  Kashiwanoha 5-1-5, Kashiwa, Chiba 277-8581, Japan\\\it
}
\begin{document}
\maketitle
\setlength{\baselineskip}{1.8em}
%
%
%
\begin{abstract}
Analytical expressions of some of the spin-spin correlation
functions up to eight lattice sites for the spin-1/2 anti-ferromagnetic 
Heisenberg chain at zero temperature without magnetic field are obtained.
The key object of our method is the generating function of two-point spin-spin correlators,
whose functional relations are derived from those for
general inhomogeneous correlation functions
previously obtained from the quantum Knizhnik-Zamolodchikov equations.
We show how the generating functions are fully determined
by their functional relations,
which leads to the two-point spin-spin correlators. 
The obtained analytical results are numerically confirmed by
the exact diagonalization for finite systems.
\end{abstract}
%
%
%
%
%
%
%
%
%
%
\newpage
\noindent
%
%
\section{Introduction}
The spin-1/2 antiferromagnetic  Heisenberg \xxx chain is one of the most
intensively studied models for one-dimensional quantum magnetism,
which is given by the Hamiltonian
\begin{align}
\mathcal{H}=\sum_{j=-\infty}^{\infty} 
\(
S_{j}^x S_{j+1}^x + S_{j}^y S_{j+1}^y + S_{j}^z S_{j+1}^z
\),
\label{Hamiltonian}
\end{align}
where ${S_j^{\alpha} = \sigma_j^{\alpha}/2}$ with  ${\sigma_j^{\alpha}}$ being the 
Pauli matrices acting on the ${j}$-th site.
Many physical quantities of this model in the thermodynamic limit
have been exactly evaluated \cite{TakaBook} by means of Bethe Ansatz \cite{Bethe31}.

The exact calculation of the correlation functions, 
on the contrary, 
still involves essential difficulties.
Among them especially significant are the two-point spin-spin correlators $\zcor{j}{j+k}$,
for which only the first- and second-neighbor correlators ($k=1,2$) had been
known until quite recently: 
\begin{align}
\zcor{j}{j+1}
& = \frac{1}{12} - \frac{1}{3} \zeta_a(1)
  = -0.147715726853315 \cdots , 
\label{first_neighbor}
\\
\zcor{j}{j+2}
& = \frac{1}{12} - \frac{4}{3} \zeta_a(1) + \zeta_a(3)
  =  0.060679769956435 \cdots , 
\label{second_neighbor}
\end{align}
where $\zeta_a(s)$ is the alternating zeta function defined by 
${\zeta_a(s) \equiv \sum_{n=1}^{\infty} (-1)^{n-1}/n^s}$, 
which is related to Riemann zeta function 
${\zeta(s) \equiv \sum_{n=1}^{\infty} 1/n^s}$ 
as
${\zeta_a(s) = (1-2^{1-s})\zeta(s)}$. 
Note that the alternating zeta function is regular at ${s=1}$ 
and is given by ${\zeta_a(1)}=\ln 2$. 

The nearest-neighbor correlator (\ref{first_neighbor}) is derived 
directly from the ground state energy which is obtained
by Hulth\'{e}n in 1938 \cite{Hulthen38}.
The first nontrivial result (\ref{second_neighbor})
was derived by Takahashi in 1977 via the strong coupling expansion 
of the ground state energy for the half-filled Hubbard model \cite{Taka77}. 
Note also that another derivation of the second-neighbor correlator 
(\ref{second_neighbor}) was given by Dittrich and 
Inozemtsev in 1997 \cite{Dittrich}.
These method, however, can not be generalized to calculate 
further correlators $\zcor{j}{j+k} \,\, (k \geq 3)$, unfortunately. 

On the other hand, 
utilizing the representation theory of the quantum affine algebra
$U_q(\hat{sl_2})$, 
Kyoto Group (Jimbo, Miki, Miwa, Nakayashiki) 
derived a multiple integral representation for arbitrary
correlation functions of the massive $XXZ$ antiferromagnetic chain
in 1992 \cite{Jimbo92, JMBook},
which is before long extended to the $XXX$ case \cite{Nakayashiki94} 
and the massless $XXZ$ case \cite{Jimbo96}. 
Furthermore the same integral representations were reproduced by 
Kitanine, Maillet, Terras \cite{Kitanine00} in the framework of 
Quantum Inverse Scattering Method. 
The integral formula for the correlation function on $n$ lattice sites 
consists of $n$-dimensional integrals, 
of which the exact evaluation, 
even for the second-neighbor correlator (\ref{second_neighbor}), 
was not succeeded for a long time. 

It is remarkable, therefore, that 
Boos and Korepin devised a method to evaluate these multiple integrals 
for $XXX$ chain in 2001
\cite{Boos01, Boos01n2}. 
It was shown that the integrand in the multiple integral formula 
can be reduced to a {\it canonical} form, 
which allows us to implement the integration.
This method was at first applied to 
a special correlation function called the Emptiness Formation Probability 
(EFP) \cite{Korepin94} which is the probability to find a ferromagnetic string 
of length $n$: 
\begin{align}
P(n) \equiv \left\bra \prod_{j=1}^{n} \( \frac{1}{2} +S_j^z \) \right\ket.
\label{Def_EFP}
\end{align}
The analytical forms of the EFP was obtained up to $n=5$ 
\cite{Boos01, Boos01n2, Boos02} 
by performing the integration: 
\begin{align}
P(2)=&
\frac{1}{3}-\frac{1}{3}\zeta_a(1)
=0.102284273146684\cdots, \\
P(3)=&
\frac{1}{4}-\zeta_a(1)+\frac{1}{2}\zeta_a(3)
=0.00762415812490254\cdots, \\
P(4)=&
\frac{1}{5}
-2\zeta_a(1)
+\frac{173}{45}\zeta_a(3)
-\frac{22}{9}\zeta_a(1)\zeta_a(3)
-\frac{17}{15}\zeta_a(3)^2
-\frac{22}{9}\zeta_a(5)
+\frac{34}{9}\zeta_a(1)\zeta_a(5) \nn\\
=& 0.000206270046519527 \cdots, \\
P(5)=&
\frac{1}{6}
- \frac{10}{3} \zeta_a(1)
+ \frac{281}{18} \zeta_a(3)
- 30 \zeta_a(1) \zeta_a(3)
- \frac{163}{3} \zeta_a(3)^2 
- \frac{1355}{36} \zeta_a(5) 
+ \frac{1960}{9} \zeta_a(1) \zeta_a(5) \nn\\
&
- \frac{85}{9} \zeta_a(3) \zeta_a(5) 
- \frac{485}{9} \zeta_a(5)^2 
+ \frac{889}{36} \zeta_a(7) 
- \frac{1645}{9} \zeta_a(1) \zeta_a(7) 
+ \frac{679}{6} \zeta_a(3) \zeta_a(7) \nn\\
=& 2.01172595898884 \cdots \times 10^{-6}. 
\label{P5}
\end{align}
Here we note that from these results the first and second-neighbor 
correlators are reproduced through the relations 
$P(2)=\zcor{j}{j+1}+\frac{1}{4}$, 
$P(3)=\zcor{j}{j+1}+\frac{1}{2}\zcor{j}{j+2}+\frac{1}{8}$,
while the third-neighbor correlator $\zcor{j}{j+3}$ can not be determined 
solely from $P(4)$. 
Applying the Boos-Korepin method to other correlation functions on four lattice sites, 
in 2003, Sakai, Shiroishi, Nishiyama and Takahashi obtained all the correlation functions 
on four lattice sites \cite{Sakai03}. 
Especially they obtained the analytical form of the third-neighbor correlator 
\begin{align}
\zcor{j}{j+3}=&
 \frac{1}{12} 
- 3 \zeta_a(1) 
+ \frac{74}{9} \zeta_a(3) 
- \frac{56}{9} \zeta_a(1) \zeta_a(3) 
- \frac{8}{3} \zeta_a(3)^2
- \frac{50}{9} \zeta_a(5) 
+ \frac{80}{9} \zeta_a(1) \zeta_a(5) \nn\\
=& -0.0502486272572352\cdots. 
\label{third_neighbor}
\end{align}
In addition, their results were extended to $XXZ$ chain 
both in massless and massive regime \cite{GoKato1, GoKato2, GoKato3}. 

In principle, 
multiple integrals for any correlation functions 
can be performed by means of Boos-Korepin method. 
However, 
$P(5)$ is the only correlation function which was calculated by this method 
on five lattice sites. 
The problem lies in the fact that 
it requires enormously intricate calculations to reduce the integrand to canonical form 
as the integral dimension increases. 

In order to proceed to further correlators,
the alternative method to calculate the EFP was formulated by 
Boos, Korepin and Smirnov in 2003 \cite{Boos03}. 
They considered the inhomogeneous \xxx model, 
in which each site carries an inhomogeneous parameter $\lam_j$. 
The homogeneous $XXX$ model corresponds to the case with 
all the inhomogeneous parameters $\lam_j$ set to be $0$.
Inhomogeneous correlation functions on $n$ lattice sites are 
considered to be functions of variables $\lams{1}{n}$. 
Below let us give a brief survey of Boos-Korepin-Smirnov approach. 
First they evaluate the multiple integrals for the inhomogeneous EFP up to $n=4$
by Boos-Korepin method. 
The results consist of only one transcendental function, 
which is a generating function of the alternating zeta series, 
with rational functions of inhomogeneous parameters $\lams{1}{n}$ as coefficients. 
From this observation they suggest an ansatz for the form of 
the general inhomogeneous correlation functions 
(the proof of which is now given \cite{Takeyama1, Takeyama2, Takeyama3}). 
Moreover they derive the functional relations for the inhomogeneous EFP 
by investigating the quantum Knizhnik-Zamolodchikov (qKZ) equations 
\cite{qkz, Frenkel, Smirnov1, Smirnov2}, 
the solutions to which are connected with 
the inhomogeneous correlation functions \cite{JMBook, Nakayashiki94, Jimbo96}. 
Together with the ansatz for the final form, 
they have obtained the explicit form for the inhomogeneous EFP up to $n=6$, 
which gives a new result of $P(6)$ in the homogeneous limit 
$\hls{1}{6}$ 
\begin{align}
P(6) = & 
  \frac{1}{7}
- 5 \zeta_a(1)
+ 46 \zeta_a(3)
- \frac{560}{3} \zeta_a(1) \zeta_a(3)
- \frac{4112}{5} \zeta_a(3)^2
- \frac{38464}{135} \zeta_a(3)^3
- \frac{9244}{35} \zeta_a(5) \nn\\
& 
+\frac{52844}{15}\zeta_a(1)\zeta_a(5)
+\frac{494}{15}\zeta_a(3)\zeta_a(5)
+\frac{76928}{45}\zeta_a(1)\zeta_a(3)\zeta_a(5)
-\frac{23332}{45}\zeta_a(3)^2\zeta_a(5) \nn\\
& 
-\frac{751592}{63}\zeta_a(5)^2
+\frac{93328}{9}\zeta_a(1)\zeta_a(5)^2
+\frac{17726}{9}\zeta_a(3)\zeta_a(5)^2
+\frac{188630}{189}\zeta_a(5)^3
+\frac{3242}{5}\zeta_a(7) \nn\\
& 
-\frac{37688}{3}\zeta_a(1)\zeta_a(7)
+\frac{133386}{5}\zeta_a(3)\zeta_a(7)
-\frac{163324}{9}\zeta_a(1)\zeta_a(3)\zeta_a(7)
-\frac{248164}{45}\zeta_a(3)^2\zeta_a(7) \nn\\
& 
+\frac{33709}{9}\zeta_a(5)\zeta_a(7)
-\frac{37726}{9}\zeta_a(3)\zeta_a(5)\zeta_a(7)
-\frac{92141}{9}\zeta_a(7)^2
+\frac{132041}{9}\zeta_a(1)\zeta_a(7)^2 \nn\\
& 
-426\zeta_a(9)
+\frac{46192}{5}\zeta_a(1)\zeta_a(9)
-\frac{375406}{15}\zeta_a(3)\zeta_a(9)
+\frac{248164}{15}\zeta_a(1)\zeta_a(3)\zeta_a(9) \nn\\
& 
+\frac{37726}{5}\zeta_a(3)^2\zeta_a(9)
+\frac{52652}{3}\zeta_a(5)\zeta_a(9)
-\frac{75452}{3}\zeta_a(1)\zeta_a(5)\zeta_a(9) \nn\\
=& 7.06812753309203 \cdots \times 10^{-9}. 
\label{P6}
\end{align}

Recently the above method was generalized to the arbitrary correlation functions 
by Boos, Shiroishi and Takahashi \cite{Boos05}. 
They obtained all the correlation functions on five lattice sites, 
especially the fourth-neighbor correlator: 
\begin{align}
\zcor{j}{j+4}
& = 
 \frac{1}{12} 
- \frac{16}{3} \zeta_a(1) 
+ \frac{290}{9} \zeta_a(3) 
- 72 \zeta_a(1) \zeta_a(3) 
- \frac{1172}{9} \zeta_a(3)^2 
- \frac{700}{9} \zeta_a(5) 
- \frac{400}{3} \zeta_a(5)^2 \nn\\
& 
+ \frac{4640}{9} \zeta_a(1) \zeta_a(5) 
- \frac{220}{9} \zeta_a(3) \zeta_a(5) 
+ \frac{455}{9} \zeta_a(7) 
- \frac{3920}{9} \zeta_a(1) \zeta_a(7) 
+ 280 \zeta_a(3) \zeta_a(7)  \nn \\
=& 0.0346527769827281 \cdots. 
\label{fourth_neighbor}
\end{align}
The calculation of general correlation functions is much complicated 
than that of the EFP, 
since only the EFP has the particular symmetry property 
$P_n(\lams{1}{n})=P_n(\lams{\sigma(1)}{\sigma(n)})$, 
where $\sigma$ denotes any element of the symmetric group $S_n$. 
In this perspective, 
we have presented a new method in the previous work \cite{SS05} 
to calculate spin-spin correlators $\zcor{j}{j+k}$, 
the central object of which is the generating function 
\cite{Izergin85, Essler96, Kitanine02, Kitanine05, Kitanine05n2}. 
The generating function has the same symmetric property as the EFP, 
which gives us a more efficient way 
to calculate spin-spin correlators $\zcor{j}{j+k}$. 
Actually the fifth-neighbor correlator has been calculated as follows: 
%
%
\begin{align}
&\zcor{j}{j+5}=
\nn\\&
\frac{1}{12}
-\frac{25}{3}\zeta_a(1)
+\frac{800}{9}\zeta_a(3)
-\frac{1192}{3}\zeta_a(1)\zeta_a(3)
-\frac{15368}{9}\zeta_a(3)^2
-608\zeta_a(3)^3
-\frac{4228}{9}\zeta_a(5)
\nn\\&
+\frac{64256}{9}\zeta_a(1)\zeta_a(5)
-\frac{976}{9}\zeta_a(3)\zeta_a(5)
+3648\zeta_a(1)\zeta_a(3)\zeta_a(5)
-\frac{3328}{3}\zeta_a(3)^2\zeta_a(5)
-\frac{76640}{3}\zeta_a(5)^2
\nn\\&
+\frac{66560}{3}\zeta_a(1)\zeta_a(5)^2
+\frac{12640}{3}\zeta_a(3)\zeta_a(5)^2
+\frac{6400}{3}\zeta_a(5)^3
+\frac{9674}{9}\zeta_a(7)
-\frac{225848}{9}\zeta_a(1)\zeta_a(7)
\nn\\&
+56952\zeta_a(3)\zeta_a(7)
-\frac{116480}{3}\zeta_a(1)\zeta_a(3)\zeta_a(7)
-\frac{35392}{3}\zeta_a(3)^2\zeta_a(7)
+7840\zeta_a(5)\zeta_a(7)
\nn\\&
-8960\zeta_a(3)\zeta_a(5)\zeta_a(7)
-\frac{66640}{3}\zeta_a(7)^2
+31360\zeta_a(1)\zeta_a(7)^2
-686\zeta_a(9)
+18368\zeta_a(1)\zeta_a(9)
\nn\\&
-53312\zeta_a(3)\zeta_a(9)
+35392\zeta_a(1)\zeta_a(3)\zeta_a(9)
+16128\zeta_a(3)^2\zeta_a(9)
\nn\\&
+38080\zeta_a(5)\zeta_a(9)
-53760\zeta_a(1)\zeta_a(5)\zeta_a(9)
\nn\\&
=-0.0308903666476093\cdots. 
\label{fifth_neighbor}
\end{align}
The aim of this paper is to show detailed calculation of the generating functions 
and further results for correlation functions. 
Especially we have obtained the analytic expressions of the spin-spin correlators 
up to $n=8$.

\section{Generating function and its functional relations}
\setcounter{equation}{0}

In this section, 
we introduce the generating function of two-point spin-spin correlators 
and derive its functional relations. 
%
\subsection{Generationg function}
First of all we define the generating function as
\begin{align}
\Pk{n}(\lams{1}{n})
\equiv
\left\bra
\prod^n_{j=1}
\left\{
\(\frac{1}{2}+S^z_j\)
+\kp\(\frac{1}{2}-S^z_j\)
\right\}
\right\ket
(\lams{1}{n}),
\label{defGF}
\end{align}
where $\kp$ is a parameter. 
We consider the inhomogeneous $XXX$ model, 
as a result of which the correlation functions on adjacent $n$ lattice sites 
depend on the inhomogeneous parameters $\lams{1}{n}$. 
From the generating function above, 
the two-point spin-spin correlators are obtained through the relation \cite{Kitanine05}
\begin{align}
\zcor{1}{n}(\lams{1}{n})
& = 
\frac{1}{2} \frac{\partial^2}{\partial \kp^2} 
\Big\{
\Pk{n}(\lams{1}{n}) 
-\Pk{n-1}(\lams{1}{n-1}) \nn\\
& 
- \Pk{n-1}(\lams{2}{n})
+\Pk{n-2}(\lams{2}{n-1})
\Big\}
\Bigg|_{\kp=1} 
- \frac{1}{4}.
\end{align}
Note that ${\Pk{n}(\lams{1}{n})}$ is a natural generalization of the EFP as 
\begin{align}
& P_n^{\kappa=0}(\lams{1}{n})
= P_n(\lams{1}{n}),
\end{align}
with other interesting relations 
\begin{align}
& P_n^{\kappa=1}(\lams{1}{n})= 1, \\
& P_n^{\kappa=-1}(\lams{1}{n})= 
\zzcor{n}(\lams{1}{n}),
\end{align}
where we denote the inhomogeneous EFP as 
$P_n(\lams{1}{n})$.
Especially by taking the homogeneous limit 
$\hls{1}{n}$, 
we have
\begin{align}
\zcor{1}{n}
& = 
\frac{1}{2} \frac{\partial^2}{\partial \kp^2} 
\Big\{
\Pk{n} -2\Pk{n-1}+\Pk{n-2}
\Big\}
\Bigg|_{\kp=1} 
- \frac{1}{4}, 
\label{zcor}
\\
& P_n^{\kappa=0} = P(n), 
\label{pn}
\\
& P_n^{\kappa=1}= 1, \\
& P_n^{\kappa=-1}= 
\zzcor{n}. 
\label{zzcor}
\end{align}

%
\subsection{Functional relations}
In preparation for deriving the functional relations for the generating function, 
first we list those for general correlation functions \cite{Boos05}, 
which follow from the qKZ equations (see also \cite{Takeyama1}): 
%
%
\begin{itemize}
\item{Translational invariance}
\begin{align}
P^{\epDs{1}{n}}_{\eps{1}{n}}(\lam_1 + \lam, \ldots, \lam_n + \lam)
=
P^{\epDs{1}{n}}_{\eps{1}{n}}(\lams{1}{n}),
\label{Trans_a}
\end{align}
\item{Transposition, Negating and Reverse order relations}
\begin{align}
&P^{\epDs{1}{n}}_{\eps{1}{n}}(\lams{1}{n})
=
P^{\eps{1}{n}}_{\epDs{1}{n}}(\Mlams{1}{n}) \nn\\
&=
P^{\MepDs{1}{n}}_{\Meps{1}{n}}(\lams{1}{n})
=
P^{\epDs{n}{1}}_{\eps{n}{1}}(\Mlams{n}{1})
\label{TNR}
\end{align}
\item{Intertwining relation}
\begin{align}
&\bar{R}
^{\epD{j}\epD{j+1}}
_{\TepD{j}\TepD{j+1}}
(\lam_j-\lam_{j+1})
P
^{\ldots,\TepD{j+1},\TepD{j},\ldots}
_{\ldots,\ep{j+1},\ep{j},\ldots}
(\ldots, \lam_{j+1},\lam_j,\ldots) \nn\\
& = 
P
^{\ldots,\epD{j},\epD{j+1},\ldots}
_{\ldots,\Tep{j},\Tep{j+1},\ldots}
(\ldots,\lam_j ,\lam_{j+1},\ldots)
\bar{R}
^{\Tep{j}\Tep{j+1}}
_{\ep{j}\ep{j+1}}
(\lam_j-\lam_{j+1})
\label{Intertwining}
\end{align}
\item{First recurrent relation}
\begin{align}
&P
^{\epD{1}\epDs{2}{n}}
_{\ep{1}\eps{2}{n}}
(\lam+1,\lam,\lams{3}{n})
=
-\delta_{\ep{1},-\ep{2}}
\epD{1}\ep{2}
P
^{\epD{2}\epDs{3}{n}}
_{-\epD{1}\eps{3}{n}}
(\lam,\lams{3}{n}) \nn\\
&P
^{\epD{1}\epDs{2}{n}}
_{\ep{1}\eps{2}{n}}
(\lam-1,\lam,\lams{3}{n})
=
-\delta_{\epD{1},-\epD{2}}
\ep{1}\epD{2}
P
^{-\ep{1}\epDs{3}{n}}
_{\ep{2}\eps{3}{n}}
(\lam,\lams{3}{n})
\label{Recurrent1_a}
\end{align}
\item{Second recurrent relation}
\begin{align}
\lim_{\il{j}}
P
^{\epDs{1}{j},\ldots,\epD{n}}
_{\eps{1}{j},\ldots,\ep{n}}
(\lams{1}{j},\ldots,\lam_n)
=
\delta_{\ep{j},\epD{j}}
\frac{1}{2}
P
^{\epDs{1}{j-1},\epDs{j+1}{n}}
_{\eps{1}{j-1},\eps{j+1}{n}}
(\lams{1}{j-1},\lams{j+1}{n})
\label{Recurrent2_a}
\end{align}
\item{Identity relation}
\begin{align}
\sum_{\eps{1}{n}}
&P^{\epDs{1}{n}}_{\eps{1}{n}}(\lams{1}{n})
=
\sum_{\epDs{1}{n}}
P^{\epDs{1}{n}}_{\eps{1}{n}}(\lams{1}{n}) \nn\\
&
=P^{+,\ldots,+}_{+,\ldots,+}(\lams{1}{n})
=P^{-,\ldots,-}_{-,\ldots,-}(\lams{1}{n})
\label{Identity}
\end{align}
\item{Reduction relation}
\begin{align}
&
P^{+,\epDs{2}{n}}_{+,\eps{2}{n}}(\lam_1,\lams{2}{n})
+
P^{-,\epDs{2}{n}}_{-,\eps{2}{n}}(\lam_1,\lams{2}{n})
=
P^{\epDs{2}{n}}_{\eps{2}{n}}(\lams{2}{n}) \nn\\
&
P^{\epDs{1}{n-1},+}_{\eps{1}{n-1},+}(\lams{1}{n-1},\lam_n)
+
P^{\epDs{1}{n-1},-}_{\eps{1}{n-1},-}(\lams{1}{n-1},\lam_n)
=
P^{\epDs{1}{n-1}}_{\eps{1}{n-1}}(\lams{1}{n-1})
\label{Reduction}
\end{align}
\end{itemize}
%
%
Here we have used the notation
\begin{align}
P^{\epDs{1}{n}}_{\eps{1}{n}}
\equiv
\bra E_1^{\epD{1}\ep{1}} \cdots E_n^{\epD{n}\ep{n}} \ket, 
\label{defP}
\end{align}
where
$E_j^{\epD{j}\ep{j}}$ is the $2 \times 2$ elementary matrix 
$(\delta_{a,\epD{j}}\delta_{b,\ep{j}})_{ab}$ acting on the $j$-th site.
$\bar{R}$ denotes an $R$-matrix of the $XXX$ model:
\begin{align}
\bar{R}(\lam)=
\begin{pmatrix}
1 &           0           &          0            & 0 \\
0 & \frac{\lam}{\lam + 1} & \frac{1}{\lam + 1}    & 0 \\
0 & \frac{1}{\lam + 1}    & \frac{\lam}{\lam + 1} & 0 \\
0 &           0           &          0            & 1 \\
\end{pmatrix}.
\label{defR}
\end{align}

Below we explore the functional relations for the generating function
$\Pk{n}(\lams{1}{n})$. 
For convenience we introduce a new function $P_{n,s}(\lams{1}{n})$ defined by 
\begin{align}
&P_{n,s}(\lams{1}{n})
\equiv
\sum_{1 \leq j_1 < \cdots < j_s \leq n}
P^{(n)}_{j_1,j_2,\ldots,j_s}(\lams{1}{n}), \nn\\
&P^{(n)}_{j_1,j_2,\ldots,j_s}(\lams{1}{n})
\equiv
P^{\eps{1}{n}}_{\eps{1}{n}}(\lams{1}{n}), \nn\\
&J=\{j_1,\ldots,j_s\}, \quad
\ep{j}=-1 \,\,\, (j \in J) , \quad
\ep{j}=1 \,\,\, (j \notin J) , 
\label{Pns}
\end{align}
which are explicitly written as, for example, 
\begin{align}
P_{4,2}(\lams{1}{4})=&
P^{(4)}_{1,2}(\lams{1}{4})+P^{(4)}_{1,3}(\lams{1}{4})+P^{(4)}_{1,4}(\lams{1}{4})\nn\\
&+P^{(4)}_{2,3}(\lams{1}{4})+P^{(4)}_{2,4}(\lams{1}{4})+P^{(4)}_{3,4}(\lams{1}{4})\nn\\
=&P^{--++}_{--++}(\lams{1}{4})+P^{-+-+}_{-+-+}(\lams{1}{4})+P^{-++-}_{-++-}(\lams{1}{4})\nn\\
&+P^{+--+}_{+--+}(\lams{1}{4})+P^{+-+-}_{+-+-}(\lams{1}{4})+P^{++--}_{++--}(\lams{1}{4}).
\end{align}
With this notation, the generating function $\Pk{n}(\lams{1}{n})$ is expanded 
in the form 
\begin{align}
\Pk{n}(\lams{1}{n})
=\sum_{s=0}^n \kp^s P_{n,s}(\lams{1}{n}).
\label{expansion}
\end{align}

Now we list the functional relations for $\Pk{n}(\lams{1}{n})$ and give their proof.
\begin{itemize}
\item{{\bf Translational invariance}}
\begin{align}
\Pk{n}(\lam_1 + \lam, \ldots, \lam_n + \lam)
=
\Pk{n}(\lams{1}{n})
\label{Trans}
\end{align}
{\flushleft {\bf Proof.}}
This is the direct consequence of that for general correlation functions (\ref{Trans_a}).
\item{{\bf Negating relation}}
\begin{align}
\Pk{n}(\Mlams{1}{n})
=
\Pk{n}(\lams{1}{n})
\label{Negating}
\end{align}
{\flushleft {\bf Proof.}}
This follows from the transposition relation (\ref{TNR})
\begin{align}
\nn
P^{\epDs{1}{n}}_{\eps{1}{n}}(\lams{1}{n})
=
P^{\eps{1}{n}}_{\epDs{1}{n}}(\Mlams{1}{n})
\end{align}
if we note that only the diagonal correlators are entered 
in the expansion (\ref{expansion}). 
\item{{\bf Symmtry relation}}
\begin{align}
\Pk{n}(\lams{1}{n})=\Pk{n}(\lams{\sigma(1)}{\sigma(n)}),
\label{symmetry}
\end{align} 
where $\sigma$ denotes any element of the symmetric group $S_n$. 
{\flushleft {\bf Proof.}}
It is enough to show that
\begin{align}
\Pk{n}(\ldots,\lam_{j+1},\lam_j,\ldots)=\Pk{n}(\ldots,\lam_j,\lam_{j+1},\ldots).
\label{sym0}
\end{align}
From the intertwining relaiton (\ref{Intertwining}), 
we have 
\begin{align}
&P^{\cdots,++,\cdots}_{\cdots,++,\cdots}(\ldots,\lam_{j+1},\lam_j,\ldots)
=P^{\cdots,++,\cdots}_{\cdots,++,\cdots}(\ldots,\lam_j,\lam_{j+1},\ldots), \nn\\
&P^{\cdots,--,\cdots}_{\cdots,--,\cdots}(\ldots,\lam_{j+1},\lam_j,\ldots)
=P^{\cdots,--,\cdots}_{\cdots,--,\cdots}(\ldots,\lam_j,\lam_{j+1},\ldots),
\label{sym1}
\end{align}
and further 
\begin{align}
&P^{\cdots,+-,\cdots}_{\cdots,+-,\cdots}(\ldots,\lam_{j+1},\lam_j,\ldots)
+(\lam_j-\lam_{j+1})
P^{\cdots,-+,\cdots}_{\cdots,+-,\cdots}(\ldots,\lam_{j+1},\lam_j,\ldots) \nn\\
& = P^{\cdots,+-,\cdots}_{\cdots,+-,\cdots}(\ldots,\lam_{j},\lam_{j+1},\ldots)
+ (\lam_j-\lam_{j+1})
P^{\cdots,+-,\cdots}_{\cdots,-+,\cdots}(\ldots,\lam_j,\lam_{j+1},\ldots), \nn\\
&P^{\cdots,-+,\cdots}_{\cdots,-+,\cdots}(\ldots,\lam_j,\lam_{j+1},\ldots)
+ (\lam_{j+1}-\lam_j)
P^{\cdots,+-,\cdots}_{\cdots,-+,\cdots}(\ldots,\lam_j,\lam_{j+1},\ldots) \nn\\
& = P^{\cdots,-+,\cdots}_{\cdots,-+,\cdots}(\ldots,\lam_{j+1},\lam_j,\ldots)
+ (\lam_{j+1}-\lam_j)
P^{\cdots,-+,\cdots}_{\cdots,+-,\cdots}(\ldots,\lam_{j+1},\lam_j,\ldots), 
\end{align}
which yield
\begin{align}
& P^{\cdots,+-,\cdots}_{\cdots,+-,\cdots}(\ldots,\lam_{j+1},\lam_j,\ldots)
+ P^{\cdots,-+,\cdots}_{\cdots,-+,\cdots}(\ldots,\lam_{j+1},\lam_j,\ldots) \nn\\
&=P^{\cdots,+-,\cdots}_{\cdots,+-,\cdots}(\ldots,\lam_j,\lam_{j+1},\ldots)
+ P^{\cdots,-+,\cdots}_{\cdots,-+,\cdots}(\ldots,\lam_j,\lam_{j+1},\ldots).
\label{sym2}
\end{align}
The above relations (\ref{sym1}), (\ref{sym2}) lead to
\begin{align}
P_{n,s}(\ldots,\lam_{j+1},\lam_j,\ldots)=P_{n,s}(\ldots,\lam_j,\lam_{j+1},\ldots),
\end{align}
from which we can conclude the relation (\ref{sym0}).
\item{{\bf First recurrent relation}}
\begin{align}
\Pk{n}(\lams{1}{n-1},\lam_{n-1}\pm 1)=
\kp \Pk{n-2}(\lams{1}{n-2})
\label{recurrent1}
\end{align}
{\flushleft {\bf Proof.}}
From the first recurrent relation for 
general correlation functions (\ref{Recurrent1_a}), 
we have
\begin{align}
P^{++,\eps{3}{n}}_{++,\eps{3}{n}}(\lam \pm 1,\lam,\lams{3}{n})
=P^{--,\eps{3}{n}}_{--,\eps{3}{n}}(\lam \pm 1,\lam,\lams{3}{n})=0,
\end{align}
and further
\begin{align}
&P^{+-,\eps{3}{n}}_{+-,\eps{3}{n}}(\lam \pm 1,\lam,\lams{3}{n})
+P^{-+,\eps{3}{n}}_{-+,\eps{3}{n}}(\lam \pm 1,\lam,\lams{3}{n}), \nn\\
&=P^{+,\eps{3}{n}}_{+,\eps{3}{n}}(\lam,\lams{3}{n})
+P^{-,\eps{3}{n}}_{-,\eps{3}{n}}(\lam,\lams{3}{n}),
\end{align}
which is calculated into 
\begin{align}
P^{+,\eps{3}{n}}_{+,\eps{3}{n}}(\lam,\lams{3}{n})
+P^{-,\eps{3}{n}}_{-,\eps{3}{n}}(\lam,\lams{3}{n})
=P^{\eps{3}{n}}_{\eps{3}{n}}(\lams{3}{n})
\end{align}
by the reduction relation (\ref{Reduction}). 
These relations are enough to show that 
\begin{align}
P_{n,s}(\lam \pm 1,\lam,\lams{3}{n})=
\left\{
\begin{matrix}
&            0            & \quad \text{for}  & \quad s=0,n             \\
&P_{n-2,s-1}(\lams{3}{n}) & \quad \text{for}  & \quad s=1,2,\ldots,n-1
\end{matrix}\right.
\end{align}
from which it follows that 
\begin{align}
&\Pk{n}(\lam \pm 1,\lam,\lams{3}{n})
=\sum_{s=0}^{n} \kp^s P_{n,s}(\lam \pm 1,\lam,\lams{3}{n}) \nn\\
&=\sum_{s=1}^{n-1} \kp^s P_{n-2,s-1}(\lams{3}{n})
=\kp \sum_{s=0}^{n-2} \kp^s P_{n-2,s}(\lams{3}{n})
=\kp \Pk{n-2}(\lams{3}{n}).
\end{align}
Taking into account the symmetry relation (\ref{symmetry}), 
we arrive at the first recurrent relation for $\Pk{n}(\lams{1}{n})$ (\ref{recurrent1}).
\item{{\bf Second recurrent relation}}
\begin{align}
\lim_{\il{n}}
\Pk{n}(\lams{1}{n-1},\lam_n)
=
\frac{1+\kp}{2}
\Pk{n-1}(\lams{1}{n-1})
\label{recurrent2}
\end{align}
{\flushleft {\bf Proof.}}
From the second recurrent relation for general correlation functions, 
we have
\begin{align}
\lim_{\il{n}}
P_{n,0}(\lams{1}{n-1},\lam_n)
=&\frac{1}{2}P_{n-1,0}(\lams{1}{n-1}), \nn\\
\lim_{\il{n}}
P_{n,n}(\lams{1}{n-1},\lam_n)
=&\frac{1}{2}P_{n-1,n-1}(\lams{1}{n-1}), \nn\\
\lim_{\il{n}}
P_{n,s}(\lams{1}{n-1},\lam_n)
=&\frac{1}{2}P_{n-1,s}(\lams{1}{n-1})
+\frac{1}{2}P_{n-1,s-1}(\lams{1}{n-1}), \nn\\
&\text{for} \quad s=1,2,\ldots,n-1. 
\end{align}
Then it follows that
\begin{align}
\lim_{\il{n}}
2\Pk{n}(\lams{1}{n-1},\lam_n)
&=P_{n-1,0}(\lams{1}{n-1})+\kp^n P_{n-1,n-1}(\lams{1}{n-1}) \nn\\
&+\sum_{s=1}^{n-1} \kp^s 
\left\{ P_{n-1,s}(\lams{1}{n-1})+P_{n-1,s-1}(\lams{1}{n-1}) \right\} \nn\\
&=(1+\kp)\sum_{s=1}^{n} \kp^{s-1} P_{n-1,s-1}(\lams{1}{n-1}) \nn\\
&=(1+\kp)\Pk{n-1}(\lams{1}{n-1}),
\end{align}
which finishes the proof of the second recurrent relation 
for $\Pk{n}(\lams{1}{n})$ (\ref{recurrent2}).
\end{itemize}
%
In this way, we have obtained all the functional relations needed for the calculation 
of $\Pk{n}(\lams{1}{n})$.

%
%
\section{General form of the generating function}
Next we consider the explicit form of the generating function $\Pk{n}(\lams{1}{n})$. 
Taking into account the general form of correlation functions \cite{Takeyama1} 
together with the symmetry relation (\ref{symmetry}), 
we can assume the generating function in the form 
\begin{align}
\Pk{n}&(\lams{1}{n})= \nn\\
&\sum_{l=0}^{[\frac{n}{2}]}
\left\{
\Ak{n,l}(\lamsB{n})
\prod_{j=1}^l
\omega(\lam_{2j-1}-\lam_{2j})
+\text{permutations}
\right\},
\label{GeneralForm}
\end{align}
where $\Ak{n,l}(\lamsB{n})$ are rational functions with known denominator 
and $\omega(\lam)$ is a transcendental function defined by 
\begin{align}
\omega(\lam)=\frac{1}{2}+
\sum_{k=1}^{\infty}
(-1)^k \frac{2k(\lam^2-1)}{\lam^2-k^2}.
\label{DefOmega}
\end{align}
The rational function $\Ak{n,l}(\lamsB{n})$ is explicitly written as 
\begin{align}
\Ak{n,l}&(\lamsB{n}) \nn\\
&=
\frac{
\prod_{k=1}^l\lam_{2k-1,2k}
\prod_{2l+1\leq k<j\leq n}\lam_{kj}
}
{
\prod_{1\leq k<j\leq n}\lam_{kj}
}
\,\Qk{n,l}(\lamsB{n}),
\label{AkForm}
\end{align}
where $\lam_{kj}=\lam_k-\lam_j$ and 
$\Qk{n,l}(\lamsB{n})$ is a polynomial with the same degree 
in each variables $\lams{1}{n}$ and also the same total degree 
as in the denominator of $\Ak{n,l}(\lamsB{n})$. 
The symmetry relation (\ref{symmetry}) for $\Pk{n}(\lams{1}{n})$ 
can be translated into that for $\Qk{n,l}(\lamsB{n})$
\begin{align}
\Qk{n,l}&(\lamsB{n}) \nn\\
&=\Qk{n,l}(\lam_2,\lam_1|\ldots|
\lam_{2l-1},\lam_{2l}||\lams{2l+1}{n}) \nn\\
&=\Qk{n,l}(\lam_1,\lam_2
|\ldots|\lam_{2j+1},\lam_{2j+2}|\lam_{2j-1},\lam_{2j}
|\ldots|\lam_{2l-1},\lam_{2l}||\lams{2l+1}{n}) \nn\\
&=\Qk{n,l}(\lam_1,\lam_2|\ldots|\lam_{2l-1},\lam_{2l}||
\lam_{2l+1},\ldots,\lam_{k+1},\lam_k,\ldots,\lam_n).
\label{qsymmetry}
\end{align} 

We shall also need the following properties for the function $\omega(\lam)$: 
\begin{align}
\omega(i \infty)=0, 
\quad
\omega(\lam \pm 1)=\alpha(\lam)+\gamma_{\pm}(\lam)\omega(\lam),
\end{align}
where
\begin{align}
\alpha(\lam)=\frac{3}{2}\frac{1}{\lam^2-1},
\quad
\gamma_{\pm}(\lam)=-\frac{\lam(\lam\pm 2)}{\lam^2-1}.
\end{align}
Moreover $\omega(\lam)$ is a generating function of the alternating zeta values
\begin{align}
\omega(\lam)=2\sum^{\infty}_{k=0}\lam^{2k}\{\zeta_a(2k-1)-\zeta_a(2k+1)\}.
\end{align}
Here we note that $\zeta_a(-1)=(1-2^2)\zeta(-1)=1/4$.

Our proposal is that the functional relations 
(\ref{Trans}), (\ref{Negating}), (\ref{symmetry}), 
(\ref{recurrent1}) and (\ref{recurrent2}) 
together with the ansatz (\ref{GeneralForm}) 
fully determine the polynomial $\Qk{n,l}(\lamsB{n})$ 
and therefore the generating function $\Pk{n}(\lams{1}{n})$.
We make a simple observation that 
the second recurrent relation (\ref{recurrent2}) is equivalent 
to the following recursion relation for $\Qk{n,l}$:
\begin{align}
\lim_{\il{n}}
&\frac{\Qk{n,l}(\lamsB{n})}{\lam_n^{2l}} \nn\\
&=
\frac{1+\kp}{2}\Qk{n-1,l}(\lamsB{n-1}).
\label{qrecurrent2}
\end{align}
Taking into account of the initial condition
\begin{align}
\Pk{0}=1,\quad \Pk{1}(\lam_1)=\frac{1+\kp}{2},
\end{align}
we can easily see that
\begin{align}
\Ak{n,0}(||\lams{1}{n})=\Qk{n,0}(||\lams{1}{n})=\ft{n}.
\end{align}
In the subsequent sections, we show explicit calculations 
of $\Pk{n}(\lams{1}{n})$ for $n \geq 2$. 

\section{Explicit calculations of the generating function}
\setcounter{equation}{0}

\subsection{$n=2$}
First we consider the generating function for $n=2$, which is given in the form 
\begin{align}
\Pk{2}(\lam_1,\lam_2)=\ft{2}+\Ak{2,1}(\lam_1,\lam_2||)\omega_{12},
\end{align}
where we have introduced the abbreviations 
$\omega_{jk}=\omega(\lam_{jk})$, 
$\lam_{jk}=\lam_j-\lam_k$. 
In this case, the rational function $\Ak{2,1}(\lam_1,\lam_2||)$ has no denominator 
and has only a constant numerator
\begin{align}
\Ak{2,1}(\lam_1,\lam_2||)=\Qk{2,1}.
\end{align}
The first recurrent relation (\ref{recurrent1}) gives a relation 
\begin{align}
\ft{2}-\frac{3}{2}\Qk{2,1}=\kp.
\end{align}
Solving this equation for $\Qk{2,1}$, we have 
\begin{align}
\Qk{2,1}=\frac{(1-\kp)^2}{6},
\end{align}
which already gives the final answer for $\Pk{2}(\lam_1,\lam_2)$:
\begin{align}
\Pk{2}(\lam_1,\lam_2)=\ft{2}+\frac{(1-\kp)^2}{6}\omega_{12}.
\end{align}
Note that the second recurrent relation (\ref{recurrent2}) is 
automatically satisfied. 
By taking the homogeneous limit $\hl{1},\hl{2}$, we have 
\begin{align}
\Pk{2}=\frac{1}{3}(1+\kp+\kp^2)-\frac{1}{3}(1-\kp)^2\zeta_a(1).
\end{align}
Through the relations (\ref{zcor}) and (\ref{pn}), 
we recover the results for $\zcor{j}{j+1}$ and $P(2)$. 

\subsection{$n=3$}

The generating function for $n=3$ can be written as 
\begin{align}
&\Pk{3}(\lam_1,\lam_2,\lam_3)
=\ft{3}+\Ak{3,1}(\lam_1,\lam_2||\lam_3)\omega_{12}
+\Ak{3,1}(\lam_1,\lam_3||\lam_2)\omega_{13}
+\Ak{3,1}(\lam_2,\lam_3||\lam_1)\omega_{23}, \nn\\
&\Ak{3,1}(\lam_1,\lam_2||\lam_3)
=\frac{\Qk{3,1}(\lam_1,\lam_2||\lam_3)}{\lam_{13}\lam_{23}}.
\label{pk3}
\end{align}
Taking into account the symmetry relation (\ref{qsymmetry}) and 
the negating relation (\ref{Negating}), 
we can assume the polynomial $\Qk{3,1}(\lam_1,\lam_2||\lam_3)$ in the form
\begin{align}
\Qk{3,1}(\lam_1,\lam_2||\lam_3)=
\ck{3,1}(0,0||0)+\ck{3,1}(1,1||0)\lam_1\lam_2
+\ck{3,1}(1,0||1)(\lam_1+\lam_2)\lam_3+\ck{3,1}(0,0||2)\lam_3^2,
\end{align}
where $\ck{3,1}(i_1,i_2||i_3)$ are some polynomials of the parameter $\kp$ 
with rational coefficients. 
The second recurrent relation (\ref{qrecurrent2}) gives the coefficient of $\lam_3^2$
\begin{align}
\ck{3,1}(0,0||2)=\frac{1+\kp}{2}\Qk{2,1}. 
\end{align}
From the translational invariance (\ref{Trans}), we have 
\begin{align}
\ck{3,1}(1,1||0)=-\ck{3,1}(1,0||1)=\ck{3,1}(0,0||2).
\end{align}
The first recurrent relation is equivalent to 
the following two relations for $\Ak{3,1}(\lam_1,\lam_2||\lam_3)$: 
\begin{align}
&\ft{3}+\alpha(\lam_{12})\Ak{3,1}(\lam_{1},\lam_{2}\pm 1||\lam_{2})
-\frac{3}{2}\Ak{3,1}(\lam_{2},\lam_{2}\pm 1||\lam_{1})
=\kp\(\frac{1+\kp}{2}\), \\[5pt]
&\Ak{3,1}(\lam_1,\lam_2||\lam_2\pm 1)
+\gamma_{\mp}(\lam_{12})\Ak{3,1}(\lam_1,\lam_2\pm1||\lam_2)=0.
\end{align}
Although these equations are an overdetermined linear system 
for the unkown coefficients $\ck{3,1}(i_1,i_2||i_3)$, 
we come to the unique solution
\begin{align}
\ck{3,1}(0,0||0)=\ck{3,1}(1,1||0)=-\ck{3,1}(1,0||1)=\ck{3,1}(0,0||2)
=\frac{1+\kp}{2}\Qk{2,1}, 
\end{align}
from which we have 
\begin{align}
\Qk{3,1}(\lam_1,\lam_2||\lam_3)=\frac{1+\kp}{2}(1+\lam_{13}\lam_{23})\Qk{2,1}.
\end{align}
Substituting it to the general form (\ref{pk3}) and 
taking the homogeneous limit $\hls{1}{3}$, 
we obtain the homogeneous generating function for $n=3$:
\begin{align}
\Pk{3}=
\frac{1}{4}(1+\kp)(1+\kp^2)
-(1-\kp)^2(1+\kp)\zeta_a(1)+\frac{1}{2}(1-\kp)^2(1+\kp)\zeta_a(3),
\end{align}
which reproduces the results for $\zcor{j}{j+2}$ and $P(3)$ through the 
relations (\ref{zcor}) and (\ref{pn}). 

\subsection{$n=4$}

The generating function for $n=4$ can be written in the form  
\begin{align}
&\Pk{4}(\lam_1,\lam_2,\lam_3,\lam_4)
=\ft{4} \nn\\
&
+\Ak{4,1}(\lam_1,\lam_2||\lam_3,\lam_4)\omega_{12}
+\Ak{4,1}(\lam_1,\lam_3||\lam_2,\lam_4)\omega_{13}
+\Ak{4,1}(\lam_1,\lam_4||\lam_2,\lam_3)\omega_{14} \nn\\
&
+\Ak{4,1}(\lam_2,\lam_3||\lam_1,\lam_4)\omega_{23}
+\Ak{4,1}(\lam_2,\lam_4||\lam_1,\lam_3)\omega_{24}
+\Ak{4,1}(\lam_3,\lam_4||\lam_1,\lam_2)\omega_{34} \nn\\
&
+\Ak{4,2}(\lam_1,\lam_2|\lam_3,\lam_4||)\omega_{12}\omega_{34}
+\Ak{4,2}(\lam_1,\lam_3|\lam_2,\lam_4||)\omega_{13}\omega_{24}
+\Ak{4,2}(\lam_1,\lam_4|\lam_2,\lam_3||)\omega_{14}\omega_{23}, \nn\\
&\Ak{4,1}(\lam_1,\lam_2||\lam_3,\lam_4)
=\frac{\Qk{4,1}(\lam_1,\lam_2||\lam_3,\lam_4)}
{\lam_{13}\lam_{23}\lam_{14}\lam_{24}}, \quad
\Ak{4,2}(\lam_1,\lam_2|\lam_3,\lam_4||)
=\frac{\Qk{4,2}(\lam_1,\lam_2|\lam_3,\lam_4||)}
{\lam_{13}\lam_{23}\lam_{14}\lam_{24}}.
\label{pk4}
\end{align}
We write the polynomials $\Qk{4,1}(\lam_1,\lam_2||\lam_3,\lam_4)$ and 
$\Qk{4,2}(\lam_1,\lam_2|\lam_3,\lam_4||)$ in the form 
\begin{align}
&\Qk{4,1}(\lam_1,\lam_2||\lam_3,\lam_4)
=\sum_{(i_1,\cdots,i_4)\in I_{4,1}}
\ck{4,1}(i_1,i_2||i_3,i_4)\lam_1^{i_1}\lam_2^{i_2}\lam_3^{i_3}\lam_4^{i_4},\\
&\Qk{4,2}(\lam_1,\lam_2|\lam_3,\lam_4||)
=\sum_{(i_1,\cdots,i_4)\in I_{4,2}}
\ck{4,2}(i_1,i_2|i_3,i_4||)\lam_1^{i_1}\lam_2^{i_2}\lam_3^{i_3}\lam_4^{i_4},\\
&I_{4,1}=I_{4,2}=\{(i_1,\cdots,i_4)|\,
0\leq i_1,\cdots,i_4\leq 2,\,i_1+i_2+i_3+i_4 \leq 4 \},
\end{align}
where $\ck{4,1}(i_1,i_2||i_3,i_4)$ and $\ck{4,2}(i_1,i_2|i_3,i_4||)$ 
are some polynomials of $\kp$ with rational coefficients. 
To determine the generating function is now equivalent to solving a 
linear system for $\ck{4,1}(i_1,i_2||i_3,i_4)$ and $\ck{4,2}(i_1,i_2|i_3,i_4||)$. 
The functional relations for the generating function 
except for the first recurrent relation 
can be explicitly written as the linear equations for 
$\ck{4,1}(i_1,i_2||i_3,i_4)$ and $\ck{4,2}(i_1,i_2|i_3,i_4||)$:
\begin{itemize}
\item{Translational invariance}
\begin{align}
&(i_1+1)\ck{4,1}(i_1+1,i_2||i_3,i_4)+(i_2+1)\ck{4,1}(i_1,i_2+1||i_3,i_4)\nn\\
&+(i_3+1)\ck{4,1}(i_1,i_2||i_3+1,i_4)+(i_4+1)\ck{4,1}(i_1,i_2||i_3,i_4+1)=\nn\\
&(i_1+1)\ck{4,2}(i_1+1,i_2|i_3,i_4||)+(i_2+1)\ck{4,2}(i_1,i_2+1|i_3,i_4||)\nn\\
&+(i_3+1)\ck{4,2}(i_1,i_2|i_3+1,i_4||)+(i_4+1)\ck{4,2}(i_1,i_2|i_3,i_4+1||)=0,\\
&\ck{4,1}(i_1,i_2||i_3,i_4)=\ck{4,2}(i_1,i_2|i_3,i_4||)=0 \quad 
\text{if} \quad (i_1,i_2,i_3,i_4) \notin I_{4,1}=I_{4,2}.\nn
\end{align}
\item{Negating relations}
\begin{align}
\ck{4,1}(i_1,i_2||i_3,i_4)=\ck{4,2}(i_1,i_2|i_3,i_4||)=0
\quad \text{if} \quad
i_1+i_2+i_3+i_4 \equiv 1 \,(\text{mod} \,2).
\end{align} 
\item{Symmetry relations}
\begin{align}
&\ck{4,1}(i_1,i_2||i_3,i_4)=\ck{4,1}(i_2,i_1||i_3,i_4)=\ck{4,1}(i_1,i_2||i_4,i_3),\\
&\ck{4,2}(i_1,i_2|i_3,i_4||)=\ck{4,2}(i_2,i_1|i_3,i_4||)=\ck{4,2}(i_1,i_2|i_4,i_3||)
=\ck{4,2}(i_3,i_4|i_1,i_2||).
\end{align}
\item{Second recurrent relations}
\begin{align}
&
\ck{4,1}(i_1,i_2||i_3,2)=\frac{1+\kp}{2}\ck{3,1}(i_1,i_2||i_3).
\end{align}
\end{itemize}

The first recurrent relation is calculated into the following four equations 
for $\Ak{4,1}$ and $\Ak{4,2}$:
\begin{align}
&
\ft{4}
+\al{13}\Ak{4,1}(\lam_{3}\pm 1,\lam_{1}||\lam_{2},\lam_{3})
+\al{23}\Ak{4,1}(\lam_{3}\pm 1,\lam_{2}||\lam_{3},\lam_{1})\nn\\
&\quad\quad\quad
-\frac{3}{2}\Ak{4,1}(\lam_{3},\lam_{3}\pm 1||\lam_{1},\lam_{2})
=
\kp\ft{2},\nn\\
&
\Ak{4,1}(\lam_{1},\lam_{2}||\lam_{3},\lam_{3}\pm 1)
-\frac{3}{2}\Ak{4,2}(\lam_{1},\lam_{2}|\lam_{3},\lam_{3}\pm 1||)
=
\kp\Ak{2,1}(\lam_{1},\lam_{2}),\nn\\
&\Ak{4,1}(\lam_{1},\lam_{3}||\lam_{2},\lam_{3}\pm 1)
+\g{13}(\lam_{2})\Ak{4,1}(\lam_{1},\lam_{3}\pm 1||\lam_{2},\lam_{3})\nn\\
&\quad\quad\quad
+\al{23}\Ak{4,2}(\lam_{1},\lam_{3}|\lam_{2},\lam_{3}\pm 1||)=0,\nn\\
&\g{23}\Ak{4,2}(\lam_{1},\lam_{3}|\lam_{2},\lam_{3}\pm 1||)
+\g{13}\Ak{4,2}(\lam_{1},\lam_{3}\pm 1|\lam_{2},\lam_{3}||)=0.
\end{align}
Solving this overdetermined linear system for the unknown coefficients 
$\ck{4,1}(i_1,i_2||i_3,i_4)$ and $\ck{4,2}(i_1,i_2|i_3,i_4||)$, 
we come to the unique solution
\begin{align}
&\Qk{4,1}(\lam_1,\lam_2||\lam_3,\lam_4) 
= 
\frac{1+\kp}{2}(1+\lam_{14} \lam_{24})\Qk{3,1}(\lam_1,\lam_2||\lam_3)
-\frac{(1-\kappa)^4}{120}(\lam_{12}^2-4), \nn\\
&\Qk{4,2}(\lam_1,\lam_2|\lam_3,\lam_4||) 
=
\frac{1}{24}
\(1+\frac{\kp}{3}\)(1+3\kp)(1-\kp)^2 
+\frac{(1-\kp)^4}{90}\(\lam_{12}^2-\frac{3}{2}\)
\(\lam_{34}^2-\frac{3}{2}\)\nn\\
&
+\frac{(1-\kp)^2}{36}
\left\{
(1+\kp)^2(\lam_{13}\lam_{24}+\lam_{14}\lam_{23}) 
+(1-\kp)^2(1+\lam_{13}\lam_{24}\lam_{14}\lam_{23})
\right\}
.
\end{align}
Taking the homogeneous limit $\hls{1}{4}$ in (\ref{pk4}), 
we obtain the homogeneous generating function for $n=4$:
\begin{align}
\Pk{4}&=
\frac{1}{5}(1+\kp+\kp^2+\kp^3+\kp^4)
+(\kp-1)^2
\left\{
-\frac{2}{3}(3+4\kp+3\kp^2)\zeta_a(1)
\right.\nn\\
&
+\frac{1}{45}(173+114\kp+173\kp^2)\zeta_a(3)
-\frac{2}{9}(11+6\kp+11\kp^2)\zeta_a(1)\zeta_a(3)
-\frac{1}{15}(17+6\kp+17\kp^2)\zeta_a(3)^2
\nn\\
&\left.
-\frac{2}{9}(11+3\kp+11\kp^2)\zeta_a(5)
+\frac{2}{9}(17+6\kp+17\kp^2)\zeta_a(1)\zeta_a(5))
\right\},
\end{align}
which reproduces the results for $\zcor{j}{j+3}$ and $P(4)$ through the 
relations (\ref{zcor}) and (\ref{pn}). 
In this case, we have also a four-spin correlation function \cite{Sakai03} 
through the relation (\ref{zzcor}):
\begin{align}
P^{\kp=-1}_4& = 2^4 \bra S^z_j S^z_{j+1} S^z_{j+2} S^z_{j+3} \ket \nn\\
&=\frac{1}{5}
-\frac{16}{3}\zeta_a(1)
+\frac{928}{45}\zeta_a(3)
-\frac{128}{9}\zeta_a(1)\zeta_a(3)
-\frac{112}{15}\zeta_a(3)^2
-\frac{152}{9}\zeta_a(5)
+\frac{224}{9}\zeta_a(1)\zeta_a(5) \nn\\
&=0.491445392361552\cdots.
\end{align}
%

\subsection{$ n \geq 5 $}

We have applied the same procedure to obtain the explicit form 
of the generating function up to $n=8$. 
First we put the polynomials $\Qk{n,l}$ into the form
\begin{align}
\Qk{n,l}(&\lamsB{n}) \nn\\
&=\sum_{(i_1,\cdots,i_n)\in I_{n,l}}
\ck{n,l}(\isB{n})\prod_{k=1}^n\lam_k^{i_k},
\end{align}
\begin{align}
I_{n,l}=\{(i_1,\cdots,i_n)|\,&
0 \leq i_{1},\cdots,i_{2l} \leq n-2,\,\,
0 \leq i_{2l+1},\cdots,i_{n} \leq 2l,\,\nn\\
&
\sum_{k=1}^{n} i_k \leq 2l(n-l-1) \}.
\end{align}
The unknown coefficients $\ck{n,l}$ should satisfy the linear equations
\begin{itemize}
\item{Translational invariance}
\begin{align}
&\sum_{k=1}^n (i_k+1)\ck{n,l}(i_1,i_2,\cdots,i_k+1,\cdots,i_{n-1},i_n)=0,\\
&\ck{n,l}(\isB{n})=0 \quad 
\text{if} \quad (i_1,\cdots,i_n) \notin I_{n,l}.\nn
\end{align}
\item{Negating relations}
\begin{align}
\ck{n,l}(\isB{n})=0
\quad \text{if} \quad
\sum_{k=1}^n i_k \equiv 1 \,(\text{mod} \,2).
\end{align} 
\item{Symmetry relations}
\begin{align}
\ck{n,l}&(\isB{n}) \nn\\
&=\ck{n,l}(i_2,i_1|\ldots|
i_{2l-1},i_{2l}||\is{2l+1}{n}) \nn\\
&=\ck{n,l}(i_1,i_2|\ldots|i_{2j+1},i_{2j+2}|
i_{2j-1},i_{2j}|\ldots|i_{2l-1},i_{2l}||\is{2l+1}{n}) \nn\\
&=\ck{n,l}(i_1,i_2|\ldots|i_{2l-1},i_{2l}||
i_{2l+1},\ldots,i_{k+1},i_k,\ldots,i_n).
\end{align} 
\item{Second recurrent relations}
\begin{align}
\ck{n,l}(\isB{n-1},2l)=\frac{1+\kp}{2}\ck{n-1,l}(\isB{n-1}).
\end{align}
\end{itemize}

The first recurrent relation for $n=5$ is given 
in terms of the rational function $\Ak{n,l}$ as: 
\begin{align}&
\ft{5}
-\frac{3}{2}\Ak{5,1}(\lam_4,\lam_4 \pm 1||\lam_1,\lam_2,\lam_3)
\nn\\
&
+\al{14}\Ak{5,1}(\lam_4 \pm 1,\lam_1||\lam_4,\lam_2,\lam_3)
+\al{24}\Ak{5,1}(\lam_4 \pm 1,\lam_2||\lam_4,\lam_1,\lam_3)
\nn\\
&
+\al{34}\Ak{5,1}(\lam_4 \pm 1,\lam_3||\lam_4,\lam_1,\lam_2)
=\kp\ft{3},\nn
\end{align}
\begin{align}&
\Ak{5,1}(\lam_4,\lam_3||\lam_4 \pm 1,\lam_1,\lam_2)
+\g{34}\Ak{5,1}(\lam_4 \pm 1,\lam_3||\lam_4,\lam_1,\lam_2)
\nn\\
&
+\al{14}\Ak{5,2}(\lam_4,\lam_3|\lam_4 \pm 1,\lam_1||\lam_2)
+\al{24}\Ak{5,2}(\lam_4,\lam_3|\lam_4 \pm 1,\lam_2||\lam_1)=0,\nn
\end{align}
\begin{align}&
\g{34}\Ak{5,2}(\lam_4,\lam_2|\lam_4 \pm 1,\lam_3||\lam_1)
+\g{24}\Ak{5,2}(\lam_4,\lam_3|\lam_4 \pm 1,\lam_2||\lam_1)=0,\nn
\end{align}
\begin{align}&
\Ak{5,1}(\lam_1,\lam_2||\lam_4,\lam_4 \pm 1,\lam_3)
-\frac{3}{2}\Ak{5,2}(\lam_4,\lam_4 \pm 1|\lam_1,\lam_2||\lam_3)
\nn\\
&
+\al{34}\Ak{5,2}(\lam_4 \pm 1,\lam_3|\lam_1,\lam_2||\lam_4)
=\kp\Ak{3,1}(\lam_1,\lam_2||\lam_3),\nn
\end{align}
\begin{align}&
\Ak{5,2}(\lam_4,\lam_3|\lam_1,\lam_2||\lam_4 \pm 1)
+\g{34}\Ak{5,2}(\lam_4 \pm 1,\lam_3|\lam_1,\lam_2||\lam_4)=0.\nn
\end{align}
The corresponding relations for $n=6,7,8$ are given in Appendix A.

We have solved this large linear system for the unknown coefficients $\ck{n,l}$ 
up to $n=8$. 
Since the obtained results are too complicated, 
we show them explicitly only for $n=5$ case below:
\begin{align}
&\Qk{5,1}(\lam_1,\lam_2||\lam_3,\lam_4,\lam_5) 
=\frac{1+\kappa}{2}(1+\lam_{15} \lam_{25})
\Qk{4,1}(\lam_1,\lam_2||\lam_3,\lam_4)
\nn\\&
-\frac{1+\kappa}{2}\frac{(1-\kappa)^4}{120}
(\lam_{12}^2-4)(4+\lam_{13}\lam_{23}+\lam_{14}\lam_{24}),
\nn\\&
\Qk{5,2}(\lam_1,\lam_2|\lam_3,\lam_4||\lam_5) 
=\frac{1+\kappa}{2}(1+\lam_{15}\lam_{25})(1+\lam_{35}\lam_{45})
\Qk{4,2}(\lam_1,\lam_2|\lam_3,\lam_4||) 
\nn\\&
+\frac{(1-\kappa)^2}{360}(\lam_{12}^2-4)(\lam_{34}^2-4)
\left\{
(1+\kappa)(1-\kappa)^2(\lam_{14}\lam_{23}+\lam_{13}\lam_{24})
+5(1+\kappa+\kappa^2+\kappa^3)
\right\} 
\nn\\&
-\frac{\kappa(1+\kappa)(1-\kappa)^2}{36}
\left\{
10-2(\lam_{12}^2+\lam_{34}^2+\lam_{14}\lam_{23}\lam_{13}\lam_{24})
\right.
\nn\\&
\left.
+(\lam_{14}\lam_{23}+\lam_{13}\lam_{24}+2)(\lam_{15}\lam_{25}+\lam_{35}\lam_{45}-1)
\right\}.
\end{align}
The homogeneous generating function is obtained by the homogeneous limit $\hl{j}$ 
\begin{align}
\Pk{5}&=
\frac{1}{6}(1+\kappa)(1-\kappa+\kappa^2)(1+\kappa+\kappa^2)
+(1-\kappa)^2(1+\kappa)
\left\{
-\frac{5}{3}(2+\kappa+2\kappa^2)\zeta_a(1)
\right.\nn\\
&
+\frac{1}{18}(281-97\kappa+281\kappa^2)\zeta_a(3)
-\frac{10}{9}(27-16\kappa+27\kappa^2)\zeta_a(1)\zeta_a(3)
\nn\\&
-\frac{1}{9}(489-368\kappa+489\kappa^2)\zeta_a(3)^2
-\frac{5}{36}(271-222\kappa+271\kappa^2)\zeta_a(5)
\nn\\&
+\frac{40}{9}(49-38\kappa+49\kappa^2)\zeta_a(1)\zeta_a(5)
-\frac{5}{9}(17-12\kappa+17\kappa^2)\zeta_a(3)\zeta_a(5)
\nn\\&
-\frac{5}{9}(97-74\kappa+97\kappa^2)\zeta_a(5)^2
+\frac{7}{36}(127-124\kappa+127\kappa^2)\zeta_a(7)
\nn\\&\left.
-\frac{35}{9}(47-38\kappa+47\kappa^2)\zeta_a(1)\zeta_a(7)
+\frac{7}{6}(97-74\kappa+97\kappa^2)\zeta_a(3)\zeta_a(7)
\right\}. 
\label{pk5}
\end{align}
In a similar way, the homogeneous generating function for $n=6$ is obtained as:
\begin{align}
\Pk{6}=\sum_{s=0}^6 \kp^s P_{6,s}, \nn
\end{align}
where
\begin{align}
P_{6,0}&=P_{6,6}=P(6), \nn \\
P_{6,1}&=P_{6,5}=
\frac{1}{7}
+2\zeta_a(1)
-\frac{302}{5}\zeta_a(3)
+312\zeta_a(1)\zeta_a(3)
+\frac{23558}{15}\zeta_a(3)^2
+\frac{26152}{45}\zeta_a(3)^3
\nn\\&
+\frac{17349}{35}\zeta_a(5)
-\frac{102604}{15}\zeta_a(1)\zeta_a(5)
-\frac{2522}{15}\zeta_a(3)\zeta_a(5)
-\frac{52304}{15}\zeta_a(1)\zeta_a(3)\zeta_a(5)
\nn\\&
+\frac{15832}{15}\zeta_a(3)^2\zeta_a(5)
+\frac{507140}{21}\zeta_a(5)^2
-\frac{63328}{3}\zeta_a(1)\zeta_a(5)^2
-\frac{12028}{3}\zeta_a(3)\zeta_a(5)^2
\nn\\&
-\frac{128060}{63}\zeta_a(5)^3
-\frac{7041}{5}\zeta_a(7)
+\frac{77798}{3}\zeta_a(1)\zeta_a(7)
-\frac{270472}{5}\zeta_a(3)\zeta_a(7)
\nn\\&
+\frac{110824}{3}\zeta_a(1)\zeta_a(3)\zeta_a(7)
+\frac{168392}{15}\zeta_a(3)^2\zeta_a(7)
-7698\zeta_a(5)\zeta_a(7)
+\frac{25612}{3}\zeta_a(3)\zeta_a(5)\zeta_a(7)
\nn\\&
+\frac{62125}{3}\zeta_a(7)^2
-\frac{89642}{3}\zeta_a(1)\zeta_a(7)^2
+\frac{4856}{5}\zeta_a(9)
-\frac{97112}{5}\zeta_a(1)\zeta_a(9)
+51048\zeta_a(3)\zeta_a(9)
\nn\\&
-\frac{168392}{5}\zeta_a(1)\zeta_a(3)\zeta_a(9)
-\frac{76836}{5}\zeta_a(3)^2\zeta_a(9)
-35500\zeta_a(5)\zeta_a(9)
+51224\zeta_a(1)\zeta_a(5)\zeta_a(9), \nn 
\end{align}
\begin{align}
P_{6,2}&=P_{6,4}=
\frac{1}{7}
+2\zeta_a(1)
+\frac{48}{5}\zeta_a(3)
-128\zeta_a(1)\zeta_a(3)
-\frac{12848}{15}\zeta_a(3)^2
-\frac{16576}{45}\zeta_a(3)^3
\nn\\&
-\frac{1734}{7}\zeta_a(5)
+3852\zeta_a(1)\zeta_a(5)
+\frac{1094}{5}\zeta_a(3)\zeta_a(5)
+\frac{33152}{15}\zeta_a(1)\zeta_a(3)\zeta_a(5)
\nn\\&
-\frac{3324}{5}\zeta_a(3)^2\zeta_a(5)
-\frac{105288}{7}\zeta_a(5)^2
+13296\zeta_a(1)\zeta_a(5)^2
+\frac{7574}{3}\zeta_a(3)\zeta_a(5)^2
+\frac{80750}{63}\zeta_a(5)^3
\nn\\&
+\frac{4866}{5}\zeta_a(7)
-16632\zeta_a(1)\zeta_a(7)
+\frac{168974}{5}\zeta_a(3)\zeta_a(7)
-23268\zeta_a(1)\zeta_a(3)\zeta_a(7)
\nn\\&
-\frac{106036}{15}\zeta_a(3)^2\zeta_a(7)
+4923\zeta_a(5)\zeta_a(7)
-\frac{16150}{3}\zeta_a(3)\zeta_a(5)\zeta_a(7)
-\frac{38717}{3}\zeta_a(7)^2
\nn\\&
+\frac{56525}{3}\zeta_a(1)\zeta_a(7)^2
-\frac{3684}{5}\zeta_a(9)
+12912\zeta_a(1)\zeta_a(9)
-\frac{161302}{5}\zeta_a(3)\zeta_a(9)
\nn\\&
+\frac{106036}{5}\zeta_a(1)\zeta_a(3)\zeta_a(9)
+9690\zeta_a(3)^2\zeta_a(9)
+22124\zeta_a(5)\zeta_a(9)
-32300\zeta_a(1)\zeta_a(5)\zeta_a(9), \nn \\
P_{6,3}&=1- 2 \left\{P_{6,0}+P_{6,1}+P_{6,2}\right\}. 
\label{pk6}
\end{align}
Using these analytical forms of the homogeneous generating functions, 
one can obtain the fourth- and fifth-neighbor correlators 
$\zcor{j}{j+4}$, 
$\zcor{j}{j+5}$ 
from the relations (\ref{zcor}) 
and also a six-spin correlation function 
\begin{align}
&
\zzcor{6} 
= P_6^{\kappa=-1}
\nn\\&
=\frac{1}{7}
-12\zeta_a(1)
+\frac{1112}{5}\zeta_a(3)
-\frac{3776}{3}\zeta_a(1)\zeta_a(3)
-\frac{100736}{15}\zeta_a(3)^2
-\frac{352768}{135}\zeta_a(3)^3
-\frac{71656}{35}\zeta_a(5)
\nn\\&
+\frac{442496}{15}\zeta_a(1)\zeta_a(5)
+\frac{15104}{15}\zeta_a(3)\zeta_a(5)
+\frac{705536}{45}
\zeta_a(1)\zeta_a(3)\zeta_a(5)
-\frac{212992}{45}\zeta_a(3)^2\zeta_a(5)
\nn\\&
-\frac{6796736}{63}\zeta_a(5)^2
+\frac{851968}{9}\zeta_a(1)\zeta_a(5)^2
+\frac{161792}{9}\zeta_a(3)\zeta_a(5)^2
+\frac{1723520}{189}\zeta_a(5)^3
+\frac{32432}{5}\zeta_a(7)
\nn\\&
-\frac{350336}{3}\zeta_a(1)\zeta_a(7)
+241888\zeta_a(3)\zeta_a(7)
-\frac{1490944}{9}\zeta_a(1)\zeta_a(3)
\zeta_a(7)
-\frac{2265088}{45}\zeta_a(3)^2\zeta_a(7)
\nn\\&
+\frac{312064}{9}\zeta_a(5)\zeta_a(7)
-\frac{344704}{9}\zeta_a(3)\zeta_a(5)\zeta_a(7)
-\frac{833168}{9}\zeta_a(7)^2
+\frac{1206464}{9}\zeta_a(1)\zeta_a(7)^2
\nn\\&
-\frac{23256}{5}\zeta_a(9)
+\frac{443008}{5}\zeta_a(1)\zeta_a(9)
-\frac{3437248}{15}\zeta_a(3)\zeta_a(9)
+\frac{2265088}{15}\zeta_a(1)\zeta_a(3)\zeta_a(9)
\nn\\&
+\frac{344704}{5}\zeta_a(3)^2\zeta_a(9)
+\frac{476096}{3}\zeta_a(5)\zeta_a(9)
-\frac{689408}{3}\zeta_a(1)\zeta_a(5)\zeta_a(9)
\nn\\&
=-0.440301669702626\ldots. 
\end{align}
Although we have obtained the explicit forms of the generating functions up to $n=8$, 
the whole expression of which for $n\geq 7$ is too complicated to be described here. 
Therefore we only present the resulting new correlation functions 
$\zcor{j}{j+6}$,
$\zcor{j}{j+7}$,
$P(7)$,
$P(8)$,
$\bra \prod_{j=1}^8 S_j^z \ket$ in Appendix B. 

%
\section{Comparison with numerical diagonalization and asymptotic formula for ${P(n)}$}
Here let us compare our new analytical expressions for correlation functions 
with numerical results by the exact diagonalization. It was 
recently known numerical data by the exact diagonalization for finite systems 
together with higher order extrapolation produce fairly good approximated 
values for correlation functions in the infinite system \cite{Sakai03, Boos05}. 
In Table \ref{diag},  
we have listed the data corresponding to the present correlation functions 
up to the system size ${N=32}$. We have 
extrapolated the data by $c_0 + c_1/N^2  + c_2/N^3  + c_3/N^4 + c_4/N^5$ and 
estimated the values in the infinite system by ${c_0}$. %
%
\begin{table}[h]
\begin{center}
\begin{small}
\caption{Numerical data of the exact diagonalization for finite systems}
\label{diag}
\begin{tabular}
{@{\hspace{\tabcolsep}\extracolsep{\fill}}cccccccc} 
\hline
Correlators & $N=24$ & $N=26$ & $N=28$ & $N=30$ & $N=32$ & Extrap. ${(c_0)}$ & Exact \\ 
\hline
$10^{12} \cdot P(7)$ 
&2.673810&3.233661&3.747345&4.212012&4.628954&8.977128&8.930907\\
$10^{15} \cdot P(8)$ 
&0.482479&0.681443&0.888130&1.093964&1.293078&4.117505&4.057495\\
-$10^2 \cdot \zcor{j}{j+5}$ 
&3.294714&3.262686&3.237667&3.217735&3.201590&3.089099&3.089037\\
$10^2 \cdot \zcor{j}{j+6}$ 
&2.665704&2.630816&2.603674&2.582122&2.564712&2.444761&2.444674\\
-$10^2 \cdot \zcor{j}{j+7}$ 
&2.561074&2.510556&2.471627&2.440946&2.416307&2.250306&2.249822\\
-$\zzcor{6}$ 
&0.453267&0.451294&0.449743&0.448501&0.447490&0.440305&0.440302\\
$\zzcor{8}$ 
&0.429047&0.425646&0.422994&0.420883&0.419175&0.407261&0.407242\\
\hline
\end{tabular}
\end{small}
\end{center}
\end{table}
%
As we can see, 
they agree quite well with our analytical results 
with more than 4 digits accuracy except for ${P(7)}$ and ${P(8)}$, 
which may be too minute to be calculated precisely by numerical simulation. 
On the other hand, it has been argued that 
the ${P(n)}$ will exhibit Gaussian decay as ${n \to \infty}$. 
The asymptotic formula proposed in \cite{KLNS, Kitanine02n2} is 
\begin{align}
P(n) \simeq An^{-\gamma}C^{-n^2},
\label{asympFormula}
\end{align}
where the Gaussian decay factor $C$ and the power-law exponent $\gamma$ are given by
\begin{align}
&C=\frac{\Gamma^2(1/4)}{\pi\sqrt{2\pi}}=1.66925\cdots,\\
&\gamma=\frac{1}{12}.
\end{align}
Unfortunately the analytic formula for the prefactor $A$ is not known yet. 
Here, however, we can estimate it from our analytical results up to $P(8)$. 
In Table \ref{asympA}, 
we calculate the ratio $A(n)\equiv P(n)/\(n^{-\gamma}C^{-n^2}\)$ for each ${P(n)}$. 
We find ${A(n)}$ tends to converge to a constant value as ${n \to \infty}$, 
which we estimate as $A= A(\infty)=0.84126\pm0.00003$.
%
%
%
\begin{table}[h]
\begin{center}
\begin{small}
\caption{Estimation of the prefactor $A$}
\label{asympA}
\begin{tabular}
{@{\hspace{\tabcolsep}\extracolsep{\fill}}cccccccc} 
\hline
$A$(1)&$A$(2)&$A$(3)&$A$(4)&$A$(5)&$A$(6)&$A$(7)&$A$(8)\\
\hline
0.8346268&0.8413643&0.8407233&0.8413280&0.8411528&0.8413073&0.8412309&0.8412895\\
\hline
\end{tabular}
\end{small}
\end{center}
\end{table}

\begin{figure}[h]
\begin{center}
\includegraphics{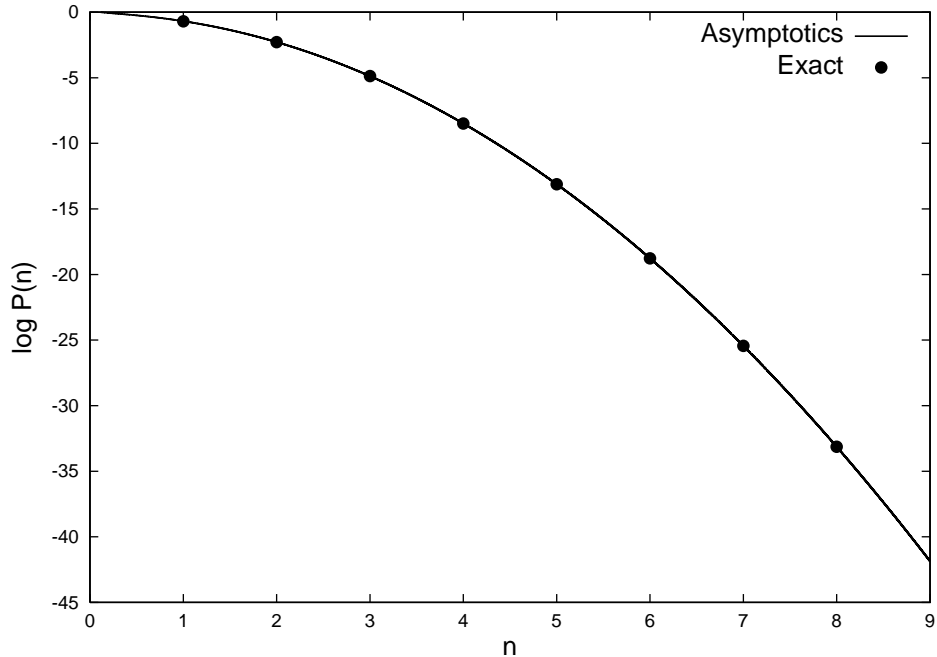}
\caption{Comparison with the asymptotic formula}
\label{asympF}
\end{center}
\end{figure}

Using the estimated value of the prefactor $A$, 
we compare our analytical results with the asymptotic formula (\ref{asympFormula}), 
which is shown in Table \ref{asympT} and Figure \ref{asympF}. 
We can conclude that our results for ${P(n})$ are well 
in accordance with the asymptotic formula (\ref{asympFormula}). 
%
%
%
\begin{table}[h]
\begin{center}
\caption{Comparison with the asymptotic formula}
\label{asympT}
\begin{tabular}
{@{\hspace{\tabcolsep}\extracolsep{\fill}}ccc} 
\hline
 $n$ &    $P(n)$ Exact     & $An^{-\gamma}C^{-n^2}$    \\
\hline
  1  & 0.5                 & 0.503974                  \\
  2  & 0.102284            & 0.102272                  \\
  3  & 7.62416 $\num{-3}$  & 7.62903 $\num{-3}$        \\
  4  & 2.06270 $\num{-4}$  & 2.06253 $\num{-4}$        \\
  5  & 2.01173 $\num{-6}$  & 2.01198 $\num{-6}$        \\
  6  & 7.06813 $\num{-9}$  & 7.06773 $\num{-9}$        \\
  7  & 8.93091 $\num{-12}$ & 8.93122 $\num{-12}$       \\
  8  & 4.05750 $\num{-15}$ & 4.05735 $\num{-15}$       \\
\hline
\end{tabular}
\end{center}
\end{table}
%
%
%
%
%
%
%
%
\section{Summary and discussion}
We introduce the generating function of the two-point spin-spin correlator 
and derive its functional relations. 
It is shown that to determine the explicit form of the generating function 
is reduced to solving a large linear system of equations 
for the unknown coefficients $\ck{n,l}$ of the polynomials $\Qk{n,l}$. 
Solving this large linear system, we obtain the generating function up to $n=8$. 
Especially for the two-point spin-spin correlators, 
we have discovered their analytical expressions 
up to the seventh-neighbor correlator $\zcor{j}{j+7}$. 
It is unrealistic, however, to apply this method to further correlation functions. 
Actually the cardinality of $I_{8,4}$, 
for instance, 
is more than $3\times10^{6}$. 
This means that in order to obtain the generating function for $n=8$, 
we had to solve the linear system with the number of unknown variables 
being the order $10^7$. 
As an alternative way to calculate the correlation functions, 
we may make use of the closed and more compact formula discovered by 
Boos, Jimbo, Miwa, Smirnov, Takeyama \cite{Takeyama1}. 
We shall investigate their formula and try to calculate further correlation
functions in the subsequent publications. 
Finally, it would be an interesting problem to generalize our approach 
by generating function to the $XXZ$ and the $XYZ$ models. 
In this respect, it is remarkable that Kitanine, Maillet, Slavnov and Terras 
have recently succeeded in calculating $\zcor{j}{j+k}$ for $\Delta=1/2$ up to $k=8$ 
from the multiple integral formula for the generating function \cite{Kitanine05n3}. 

%
\section*{Acknowledgement}
We thank T. Sakai for giving us the numerical data of the exact diagonalization. 
We are grateful to 
H. Boos, F. G\"{o}hmann, M. Jimbo, C. Korff, A. Kl\"{u}mper, 
M.J. Martins, T. Miwa, K. Sakai and Z. Tsuboi 
for valuable discussions. 
This work is in part supported by Grant-in-Aid for the Scientific Research (B) No.~14340099 
from the Ministry of Education, Culture, Sports, Science and Technology, Japan. 

%
\begin{appendix}
%
%
%
\section{First recurrent relations}
In this appendix, we write down the explicit form of the first recurrent relations 
in terms of the rational function $\Ak{n,l}$ for $n=6,7,8$.
%
\subsection{$n=6$}
\begin{align}
&
\ft{6}
-\frac{3}{2}\Ak{6,1}(\lam_5,\lam_5 \pm 1||\lam_1,\lam_2,\lam_3,\lam_4)
+\al{15}\Ak{6,1}(\lam_5 \pm 1,\lam_1||\lam_5,\lam_2,\lam_3,\lam_4)
\nn\\
&
+\al{25}\Ak{6,1}(\lam_5 \pm 1,\lam_2||\lam_5,\lam_1,\lam_3,\lam_4)
+\al{35}\Ak{6,1}(\lam_5 \pm 1,\lam_3||\lam_5,\lam_1,\lam_2,\lam_4)
\nn\\
&
+\al{45}\Ak{6,1}(\lam_5 \pm 1,\lam_4||\lam_5,\lam_1,\lam_2,\lam_3)
=\ft{4},\nn
\end{align}
\begin{align}
&
\Ak{6,1}(\lam_5,\lam_4||\lam_5 \pm 1,\lam_1,\lam_2,\lam_3)
+\g{45}\Ak{6,1}(\lam_5 \pm 1,\lam_4||\lam_5,\lam_1,\lam_2,\lam_3)
\nn\\
&
+\al{15}\Ak{6,2}(\lam_5,\lam_4|\lam_5 \pm 1,\lam_1||\lam_2,\lam_3)
+\al{25}\Ak{6,2}(\lam_5,\lam_4|\lam_5 \pm 1,\lam_2||\lam_1,\lam_3)
\nn\\
&
+\al{35}\Ak{6,2}(\lam_5,\lam_4|\lam_5 \pm 1,\lam_3||\lam_1,\lam_2)
=0,\nn
\end{align}
\begin{align}
&
\g{45}\Ak{6,2}(\lam_5,\lam_3|\lam_5 \pm 1,\lam_4||\lam_1,\lam_2)
+\g{35}\Ak{6,2}(\lam_5,\lam_4|\lam_5 \pm 1,\lam_3||\lam_1,\lam_2)
=0,
\nn
\end{align}
\begin{align}&
\Ak{6,1}(\lam_2,\lam_3||\lam_5,\lam_5 \pm 1,\lam_1,\lam_4)
-\frac{3}{2}\Ak{6,2}(\lam_5,\lam_5 \pm 1|\lam_2,\lam_3||\lam_1,\lam_4)
\nn\\
&
+\al{15}\Ak{6,2}(\lam_5 \pm 1,\lam_1|\lam_2,\lam_3||\lam_5,\lam_4)
+\al{45}\Ak{6,2}(\lam_5 \pm 1,\lam_4|\lam_2,\lam_3||\lam_5,\lam_1)
\nn\\
&
=\kp\Ak{4,1}(\lam_2,\lam_3||\lam_1,\lam_4),
\nn
\end{align}
\begin{align}
&
\Ak{6,2}(\lam_1,\lam_4|\lam_2,\lam_3||\lam_5,\lam_5 \pm 1)
-\frac{3}{2}\Ak{6,3}(\lam_5,\lam_5 \pm 1|\lam_1,\lam_4|\lam_2,\lam_3)
=\kp\Ak{4,2}(\lam_1,\lam_4|\lam_2,\lam_3),
\nn
\end{align}
\begin{align}
&
\Ak{6,2}(\lam_5,\lam_4|\lam_1,\lam_2||\lam_5 \pm 1,\lam_3)
+\g{45}\Ak{6,2}(\lam_5 \pm 1,\lam_4|\lam_1,\lam_2||\lam_5,\lam_3)
\nn\\
&
+\al{35}\Ak{6,3}(\lam_5,\lam_4|\lam_5 \pm 1,\lam_3|\lam_1,\lam_2)
=0,
\nn
\end{align}
\begin{align}
&
\g{45}\Ak{6,3}(\lam_5,\lam_3|\lam_5 \pm 1,\lam_4|\lam_1,\lam_2)
+\g{35}\Ak{6,3}(\lam_5,\lam_4|\lam_5 \pm 1,\lam_3|\lam_1,\lam_2)=0.\nn
\end{align}
%
\subsection{$n=7$}
\begin{align}
&\ft{7}
-\frac{3}{2}\Ak{7,1}(\lam_6,\lam_6 \pm 1||\lam_1,\lam_2,\lam_3,\lam_4,\lam_5)
\nn\\
&
+\al{16}\Ak{7,1}(\lam_6 \pm 1,\lam_1||\lam_6,\lam_2,\lam_3,\lam_4,\lam_5)
+\al{26}\Ak{7,1}(\lam_6 \pm 1,\lam_2||\lam_6,\lam_1,\lam_3,\lam_4,\lam_5)
\nn\\
&
+\al{36}\Ak{7,1}(\lam_6 \pm 1,\lam_3||\lam_6,\lam_1,\lam_2,\lam_4,\lam_5)
+\al{46}\Ak{7,1}(\lam_6 \pm 1,\lam_4||\lam_6,\lam_1,\lam_2,\lam_3,\lam_5)
\nn\\
&
+\al{56}\Ak{7,1}(\lam_6 \pm 1,\lam_5||\lam_6,\lam_1,\lam_2,\lam_3,\lam_4)
=\kp\ft{5}
,\nn
\end{align}
\begin{align}
&\Ak{7,1}(\lam_6,\lam_5||\lam_6 \pm 1,\lam_1,\lam_2,\lam_3,\lam_4)
+\g{56}\Ak{7,1}(\lam_6 \pm 1,\lam_5||\lam_6,\lam_1,\lam_2,\lam_3,\lam_4)
\nn\\
&
+\al{16}\Ak{7,2}(\lam_6,\lam_5|\lam_6 \pm 1,\lam_1||\lam_2,\lam_3,\lam_4)
+\al{26}\Ak{7,2}(\lam_6,\lam_5|\lam_6 \pm 1,\lam_2||\lam_1,\lam_3,\lam_4)
\nn\\
&
+\al{36}\Ak{7,2}(\lam_6,\lam_5|\lam_6 \pm 1,\lam_3||\lam_1,\lam_2,\lam_4)
+\al{46}\Ak{7,2}(\lam_6,\lam_5|\lam_6 \pm 1,\lam_4||\lam_1,\lam_2,\lam_3)
=0,
\nn
\end{align}
\begin{align}
\g{56}\Ak{7,2}(\lam_6,\lam_4|\lam_6 \pm 1,\lam_5||\lam_1,\lam_2,\lam_3)
+\g{46}\Ak{7,2}(\lam_6,\lam_5|\lam_6 \pm 1,\lam_4||\lam_1,\lam_2,\lam_3)=0,\nn
\end{align}
\begin{align}
&\Ak{7,1}(\lam_3,\lam_4||\lam_6,\lam_6 \pm 1,\lam_1,\lam_2,\lam_5)
-\frac{3}{2}\Ak{7,2}(\lam_6,\lam_6 \pm 1,\lam_3,\lam_4||\lam_1,\lam_2,\lam_5)
\nn\\
&
+\al{16}\Ak{7,2}(\lam_6 \pm 1,\lam_1|\lam_3,\lam_4||\lam_6,\lam_2,\lam_5)
+\al{26}\Ak{7,2}(\lam_6 \pm 1,\lam_2|\lam_3,\lam_4||\lam_6,\lam_1,\lam_5)
\nn\\
&
+\al{56}\Ak{7,2}(\lam_6 \pm 1,\lam_5|\lam_3,\lam_4||\lam_6,\lam_1,\lam_2)
=\kp\Ak{5,1}(\lam_3,\lam_4||\lam_1,\lam_2,\lam_5),
\nn
\end{align}
\begin{align}
&\Ak{7,2}(\lam_6,\lam_5|\lam_2,\lam_3||\lam_6 \pm 1,\lam_1,\lam_4)
+\g{56}\Ak{7,2}(\lam_6 \pm 1,\lam_5|\lam_2,\lam_3||\lam_6,\lam_1,\lam_4)
\nn\\
&
+\al{16}\Ak{7,3}(\lam_6,\lam_5|\lam_6 \pm 1,\lam_1|\lam_2,\lam_3||\lam_4)
+\al{46}\Ak{7,3}(\lam_6,\lam_5|\lam_6 \pm 1,\lam_4|\lam_2,\lam_3||\lam_1)
=0,
\nn
\end{align}
\begin{align}
\g{56}\Ak{7,3}(\lam_6,\lam_4|\lam_6 \pm 1,\lam_5|\lam_2,\lam_3||\lam_1)
+\g{46}\Ak{7,3}(\lam_6,\lam_5|\lam_6 \pm 1,\lam_4|\lam_2,\lam_3||\lam_1)
=0,\nn
\end{align}
\begin{align}
&\Ak{7,2}(\lam_1,\lam_4|\lam_2,\lam_3||\lam_6,\lam_6 \pm 1,\lam_5)
-\frac{3}{2}\Ak{7,3}(\lam_6,\lam_6 \pm 1|\lam_1,\lam_4|\lam_2,\lam_3||\lam_5)
\nn\\
&
+\al{56}\Ak{7,3}(\lam_6 \pm 1,\lam_5|\lam_1,\lam_4|\lam_2,\lam_3||\lam_6)
=\kp\Ak{5,2}(\lam_1,\lam_4|\lam_2,\lam_3||\lam_5),\nn
\end{align}
\begin{align}
\Ak{7,3}(\lam_6,\lam_5|\lam_1,\lam_4|\lam_2,\lam_3||\lam_6 \pm 1)
+\g{56}\Ak{7,3}(\lam_6 \pm 1,\lam_5|\lam_1,\lam_4|\lam_2,\lam_3||\lam_6)
=0.
\nn
\end{align}
%
\subsection{$n=8$}
\begin{align}
&
\ft{8}
-\frac{3}{2}\Ak{8,1}(\lam_7,\lam_7 \pm 1||\lam_1,\lam_2,\lam_3,\lam_4,\lam_5,\lam_6)
+\al{17}\Ak{8,1}(\lam_7 \pm 1,\lam_1||\lam_7,\lam_2,\lam_3,\lam_4,\lam_5,\lam_6)
\nn\\&
+\al{27}\Ak{8,1}(\lam_7 \pm 1,\lam_2||\lam_7,\lam_1,\lam_3,\lam_4,\lam_5,\lam_6)
+\al{37}\Ak{8,1}(\lam_7 \pm 1,\lam_3||\lam_7,\lam_1,\lam_2,\lam_4,\lam_5,\lam_6)
\nn\\&
+\al{47}\Ak{8,1}(\lam_7 \pm 1,\lam_4||\lam_7,\lam_1,\lam_2,\lam_3,\lam_5,\lam_6)
+\al{57}\Ak{8,1}(\lam_7 \pm 1,\lam_5||\lam_7,\lam_1,\lam_2,\lam_3,\lam_4,\lam_6)
\nn\\&
+\al{67}\Ak{8,1}(\lam_7 \pm 1,\lam_6||\lam_7,\lam_1,\lam_2,\lam_3,\lam_4,\lam_5)
=\kp\ft6
,\nn
\end{align}
\begin{align}
&
\Ak{8,1}(\lam_7,\lam_6||\lam_7 \pm 1,\lam_1,\lam_2,\lam_3,\lam_4,\lam_5)
+\g{67}\Ak{8,1}(\lam_7 \pm 1,\lam_6||\lam_7,\lam_1,\lam_2,\lam_3,\lam_4,\lam_5)
\nn\\&
+\al{17}\Ak{8,2}(\lam_7,\lam_6|\lam_7 \pm 1,\lam_1||\lam_2,\lam_3,\lam_4,\lam_5)
+\al{27}\Ak{8,2}(\lam_7,\lam_6|\lam_7 \pm 1,\lam_2||\lam_1,\lam_3,\lam_4,\lam_5)
\nn\\&
+\al{37}\Ak{8,2}(\lam_7,\lam_6|\lam_7 \pm 1,\lam_3||\lam_1,\lam_2,\lam_4,\lam_5)
+\al{47}\Ak{8,2}(\lam_7,\lam_6|\lam_7 \pm 1,\lam_4||\lam_1,\lam_2,\lam_3,\lam_5)
\nn\\&
+\al{57}\Ak{8,2}(\lam_7,\lam_6|\lam_7 \pm 1,\lam_5||\lam_1,\lam_2,\lam_3,\lam_4)
=0,\nn
\end{align}
\begin{align}
&
\g{67}\Ak{8,2}(\lam_7,\lam_5|\lam_7 \pm 1,\lam_6||\lam_1,\lam_2,\lam_3,\lam_4)
+\g{57}\Ak{8,2}(\lam_7,\lam_6|\lam_7 \pm 1,\lam_5||\lam_1,\lam_2,\lam_3,\lam_4)
=0,\nn
\end{align}
\begin{align}
&
\Ak{8,1}(\lam_4,\lam_5||\lam_7,\lam_7 \pm 1,\lam_1,\lam_2,\lam_3,\lam_6)
-\frac{3}{2}\Ak{8,2}(\lam_7,\lam_7 \pm 1|\lam_4,\lam_5||\lam_1,\lam_2,\lam_3,\lam_6)
\nn\\&
+\al{17}\Ak{8,2}(\lam_7 \pm 1,\lam_1|\lam_4,\lam_5||\lam_7,\lam_2,\lam_3,\lam_6)
+\al{27}\Ak{8,2}(\lam_7 \pm 1,\lam_2|\lam_4,\lam_5||\lam_7,\lam_1,\lam_3,\lam_6)
\nn\\&
+\al{37}\Ak{8,2}(\lam_7 \pm 1,\lam_3|\lam_4,\lam_5||\lam_7,\lam_1,\lam_2,\lam_6)
+\al{67}\Ak{8,2}(\lam_7 \pm 1,\lam_6|\lam_4,\lam_5||\lam_7,\lam_1,\lam_2,\lam_3)
\nn\\&
=\kp\Ak{6,1}(\lam_4,\lam_5||\lam_1,\lam_2,\lam_3,\lam_6)
,\nn
\end{align}
\begin{align}
&
\Ak{8,2}(\lam_7,\lam_6|\lam_3,\lam_4||\lam_7 \pm 1,\lam_1,\lam_2,\lam_5)
+\g{67}\Ak{8,2}(\lam_7 \pm 1,\lam_6|\lam_3,\lam_4||\lam_7,\lam_1,\lam_2,\lam_5)
\nn\\&
+\al{17}\Ak{8,3}(\lam_7,\lam_6|\lam_7 \pm 1,\lam_1|\lam_3,\lam_4||\lam_2,\lam_5)
+\al{27}\Ak{8,3}(\lam_7,\lam_6|\lam_7 \pm 1,\lam_2|\lam_3,\lam_4||\lam_1,\lam_5)
\nn\\&
+\al{57}\Ak{8,3}(\lam_7,\lam_6|\lam_7 \pm 1,\lam_5|\lam_3,\lam_4||\lam_1,\lam_2)
=0,\nn
\end{align}
\begin{align}
&
\g{67}\Ak{8,3}(\lam_7,\lam_5|\lam_7 \pm 1,\lam_6|\lam_3,\lam_4||\lam_1,\lam_2)
+\g{57}\Ak{8,3}(\lam_7,\lam_6|\lam_7 \pm 1,\lam_5|\lam_3,\lam_4||\lam_1,\lam_2)
=0,\nn
\end{align}
\begin{align}
&
\Ak{8,2}(\lam_2,\lam_5|\lam_3,\lam_4||\lam_7,\lam_7 \pm 1,\lam_1,\lam_6)
-\frac{3}{2}\Ak{8,3}(\lam_7,\lam_7 \pm 1|\lam_2,\lam_5|\lam_3,\lam_4||\lam_1,\lam_6)
\nn\\&
+\al{17}\Ak{8,3}(\lam_7 \pm 1,\lam_1|\lam_2,\lam_5|\lam_3,\lam_4||\lam_7,\lam_6)
+\al{67}\Ak{8,3}(\lam_7 \pm 1,\lam_6|\lam_2,\lam_5|\lam_3,\lam_4||\lam_7,\lam_1)
\nn\\&
=\kp\Ak{6,2}(\lam_2,\lam_5|\lam_3,\lam_4||\lam_1,\lam_6)
,\nn
\end{align}
\begin{align}
&
\Ak{8,3}(\lam_1,\lam_6|\lam_2,\lam_5|\lam_3,\lam_4||\lam_7,\lam_7 \pm 1)
-\frac{3}{2}\Ak{8,4}(\lam_7,\lam_7 \pm 1|\lam_1,\lam_6|\lam_2,\lam_5|\lam_3,\lam_4||)
=0,\nn
\end{align}
\begin{align}
&
\Ak{8,3}(\lam_7,\lam_6|\lam_1,\lam_4|\lam_2,\lam_3||\lam_7 \pm 1,\lam_5)
+\g{67}\Ak{8,3}(\lam_7 \pm 1,\lam_6|\lam_1,\lam_4|\lam_2,\lam_3||\lam_7,\lam_5)
\nn\\&
+\al{57}\Ak{8,4}(\lam_7,\lam_6|\lam_7 \pm 1,\lam_5|\lam_1,\lam_4|\lam_2,\lam_3||)
=\kp\Ak{6,3}(\lam_1,\lam_6|\lam_2,\lam_5|\lam_3,\lam_4||)
,\nn
\end{align}
\begin{align}
&
\g{67}\Ak{8,4}(\lam_7,\lam_5|\lam_7 \pm 1,\lam_6|\lam_1,\lam_4|\lam_2,\lam_3||)
+\g{57}\Ak{8,4}(\lam_7,\lam_6|\lam_7 \pm 1,\lam_5|\lam_1,\lam_4|\lam_2,\lam_3||)=0. \nn
\end{align}
%
%
%
%
\section{Exact results for the correlation functions}
%
\subsection{$n=7$}
%
\begin{align}
&P(7)=
\nn\\&
\frac{1}{8}
-7\zeta_a(1)
+\frac{1001}{9}\zeta_a(3)
-\frac{7252}{9}\zeta_a(1)\zeta_a(3)
-\frac{1561868}{225}\zeta_a(3)^2
-\frac{9451204}{675}\zeta_a(3)^3
\nn\\&
-\frac{6143}{5}\zeta_a(5)
+\frac{6923672}{225}\zeta_a(1)\zeta_a(5)
+\frac{418082}{75}\zeta_a(3)\zeta_a(5)
+\frac{18902408}{225}\zeta_a(1)\zeta_a(3)\zeta_a(5)
\nn\\&
-\frac{39912418}{675}\zeta_a(3)^2\zeta_a(5)
-\frac{126766694}{225}\zeta_a(5)^2
+\frac{317214352}{225}\zeta_a(1)\zeta_a(5)^2
\nn\\&
+\frac{585033736}{675}\zeta_a(3)\zeta_a(5)^2
+\frac{365558336}{135}\zeta_a(5)^3
+\frac{6362629}{900}\zeta_a(7)
-\frac{12336338}{45}\zeta_a(1)\zeta_a(7)
\nn\\&
+\frac{96396671}{75}\zeta_a(3)\zeta_a(7)
-\frac{118694758}{45}\zeta_a(1)\zeta_a(3)\zeta_a(7)
-\frac{1619922899}{675}\zeta_a(3)^2\zeta_a(7)
\nn\\&
+\frac{460897003}{450}\zeta_a(5)\zeta_a(7)
-\frac{235902856}{225}\zeta_a(1)\zeta_a(5)\zeta_a(7)
-\frac{2485833539}{225}\zeta_a(3)\zeta_a(5)\zeta_a(7)
\nn\\&
+\frac{149517466}{135}\zeta_a(5)^2\zeta_a(7)
-\frac{6341605627}{900}\zeta_a(7)^2
+\frac{202042386}{5}\zeta_a(1)\zeta_a(7)^2
\nn\\&
-\frac{5517330469}{675}\zeta_a(3)\zeta_a(7)^2
-\frac{353011729}{135}\zeta_a(5)\zeta_a(7)^2
-\frac{111342553}{30}\zeta_a(7)^3
\nn\\&
-\frac{85876}{5}\zeta_a(9)
+\frac{20098932}{25}\zeta_a(1)\zeta_a(9)
-\frac{1216542866}{225}\zeta_a(3)\zeta_a(9)
\nn\\&
+\frac{2569373212}{225}\zeta_a(1)\zeta_a(3)\zeta_a(9)
+\frac{1439237282}{75}\zeta_a(3)^2\zeta_a(9)
+\frac{909663307}{75}\zeta_a(5)\zeta_a(9)
\nn\\&
-\frac{15657023984}{225}\zeta_a(1)\zeta_a(5)\zeta_a(9)
+\frac{2716133644}{225}\zeta_a(3)\zeta_a(5)\zeta_a(9)
\nn\\&
+\frac{291429628}{45}\zeta_a(5)^2\zeta_a(9)
+\frac{1617984697}{450}\zeta_a(7)\zeta_a(9)
-\frac{4262009024}{225}\zeta_a(1)\zeta_a(7)\zeta_a(9)
\nn\\&
-\frac{235917262}{75}\zeta_a(3)\zeta_a(7)\zeta_a(9)
+\frac{63624316}{5}\zeta_a(5)\zeta_a(7)\zeta_a(9)
-\frac{137862718}{25}\zeta_a(9)^2
\nn\\&
+\frac{943669048}{25}\zeta_a(1)\zeta_a(9)^2
-\frac{572618844}{25}\zeta_a(3)\zeta_a(9)^2
+\frac{280731}{25}\zeta_a(11)
\nn\\&
-\frac{2798796}{5}\zeta_a(1)\zeta_a(11)
+\frac{929298293}{225}\zeta_a(3)\zeta_a(11)
-\frac{398911436}{45}\zeta_a(1)\zeta_a(3)\zeta_a(11)
\nn\\&
-\frac{3752306866}{225}\zeta_a(3)^2\zeta_a(11)
-\frac{361700449}{30}\zeta_a(5)\zeta_a(11)
+\frac{3045598864}{45}\zeta_a(1)\zeta_a(5)\zeta_a(11)
\nn\\&
-\frac{122670394}{45}\zeta_a(3)\zeta_a(5)\zeta_a(11)
-\frac{49990534}{3}\zeta_a(5)^2\zeta_a(11)
+\frac{758244949}{90}\zeta_a(7)\zeta_a(11)
\nn\\&
-\frac{2595089882}{45}\zeta_a(1)\zeta_a(7)\zeta_a(11)
+\frac{174966869}{5}\zeta_a(3)\zeta_a(7)\zeta_a(11)
\nn\\&
=8.9309068422694165026200597347158906769902357259448
\cdots\times 10^{-12} \nn
\end{align}
%
\begin{align}
&\zcor{j}{j+6}=
\nn\\&
\frac{1}{12}
-12\zeta_a(1)
+\frac{1795}{9}\zeta_a(3)
-\frac{4472}{3}\zeta_a(1)\zeta_a(3)
-\frac{178028}{15}\zeta_a(3)^2
-\frac{1066624}{45}\zeta_a(3)^3
\nn\\&
-\frac{16712}{9}\zeta_a(5)
+\frac{2292656}{45}\zeta_a(1)\zeta_a(5)
+\frac{25892}{5}\zeta_a(3)\zeta_a(5)
+\frac{2133248}{15}\zeta_a(1)\zeta_a(3)\zeta_a(5)
\nn\\&
-\frac{13641344}{135}\zeta_a(3)^2\zeta_a(5)
-\frac{42413752}{45}\zeta_a(5)^2
+\frac{106975616}{45}\zeta_a(1)\zeta_a(5)^2
\nn\\&
+\frac{21892352}{15}\zeta_a(3)\zeta_a(5)^2
+\frac{123425728}{27}\zeta_a(5)^3
+\frac{415688}{45}\zeta_a(7)
-\frac{3822280}{9}\zeta_a(1)\zeta_a(7)
\nn\\&
+\frac{96387844}{45}\zeta_a(3)\zeta_a(7)
-\frac{39846464}{9}\zeta_a(1)\zeta_a(3)\zeta_a(7)
-\frac{546452032}{135}\zeta_a(3)^2\zeta_(7)
\nn\\&
+\frac{74592812}{45}\zeta_a(5)\zeta_a(7)
-\frac{76339648}{45}\zeta_a(1)\zeta_a(5)\zeta_a(7)
-\frac{2517249056}{135}\zeta_a(3)\zeta_a(5)\zeta_a(7)
\nn\\&
+\frac{16853536}{9}\zeta_a(5)^2\zeta_a(7)
-\frac{541856504}{45}\zeta_a(7)^2
+\frac{205138304}{3}\zeta_a(1)\zeta_a(7)^2
\nn\\&
-\frac{1860007072}{135}\zeta_a(3)\zeta_a(7)^2
-\frac{118962592}{27}\zeta_a(5)\zeta_a(7)^2
-\frac{18790912}{3}\zeta_a(7)^3
\nn\\&
-\frac{61544}{3}\zeta_a(9)
+\frac{5997376}{5}\zeta_a(1)\zeta_a(9)
-\frac{399212968}{45}\zeta_a(3)\zeta_a(9)
\nn\\&
+\frac{286358912}{15}\zeta_a(1)\zeta_a(3)\zeta_a(9)
+\frac{161792512}{5}\zeta_a(3)^2\zeta_(9)
+\frac{932984528}{45}\zeta_a(5)\zeta_a(9)
\nn\\&
-\frac{5297189632}{45}\zeta_a(1)\zeta_a(5)\zeta_a(9)
+\frac{914345152}{45}\zeta_a(3)\zeta_a(5)\zeta_a(9)
\nn\\&
+\frac{98258944}{9}\zeta_a(5)^2\zeta_a9)
+\frac{274815128}{45}\zeta_a(7)\zeta_a(9)
-\frac{1439618432}{45}\zeta_a(1)\zeta_a(7)\zeta_a(9)
\nn\\&
-\frac{80018176}{15}\zeta_a(3)\zeta_a(7)\zeta_a(9)
+21475328\zeta_a(5)\zeta_a(7)\zeta_a(9)
-\frac{47617024}{5}\zeta_a(9)^2
\nn\\&
+\frac{320072704}{5}\zeta_a(1)\zeta_a(9)^2
-\frac{193277952}{5}\zeta_a(3)\zeta_a(9)^2
+\frac{194194}{15}\zeta_a(11)
\nn\\&
-824208\zeta_a(1)\zeta_a(11)
+\frac{303693544}{45}\zeta_a(3)\zeta_a(11)
-14803712\zeta_a(1)\zeta_a(3)\zeta_a(11)
\nn\\&
-\frac{1265342848}{45}\zeta_a(3)^2\zeta_a(11)
-\frac{184974328}{9}\zeta_a(5)\zeta_a(11)
\nn\\&
+\frac{1029873152}{9}\zeta_a(1)\zeta_a(5)\zeta_a(11)
-\frac{40798912}{9}\zeta_a(3)\zeta_a(5)\zeta_a(11)
\nn\\&
-\frac{84367360}{3}\zeta_a(5)^2\zeta_a(11)
+\frac{130946816}{9}\zeta_a(7)\zeta_a(11)
\nn\\&
-\frac{880199936}{9}\zeta_a(1)\zeta_a(7)\zeta_a(1)
+59057152\zeta_a(3)\zeta_a(7)\zeta_a(11)
\nn\\&
=0.024446738327958906541769539023700938400615318034987\cdots \nn
\end{align}
%
\subsection{$n=8$}
%
\begin{align}
&P(8)=
\nn\\&
\frac{1}{9}
-\frac{28}{3}\zeta_a(1)
+\frac{2114}{9}\zeta_a(3)
-2744\zeta_a(1)\zeta_a(3)
-\frac{9140908}{225}\zeta_a(3)^2
-\frac{65456384}{225}\zeta_a(3)^3
\nn\\&
-\frac{66044}{15}\zeta_a(5)
+\frac{13790504}{75}\zeta_a(1)\zeta_a(5)
+\frac{4177472}{45}\zeta_a(3)\zeta_a(5)
+\frac{130912768}{75}\zeta_a(1)\zeta_a(3)\zeta_a(5)
\nn\\&
-\frac{509355808}{225}\zeta_a(3)^2\zeta_a(5)
-\frac{2115314412}{175}\zeta_a(5)^2
+\frac{4575869168}{75}\zeta_a(1)\zeta_a(5)^2
\nn\\&
+\frac{6207314008}{75}\zeta_a(3)\zeta_a(5)^2
+\frac{1229529472}{75}\zeta_a(3)^2\zeta_a(5)^2
+\frac{117255771032}{175}\zeta_a(5)^3
\nn\\&
-\frac{2459058944}{25}\zeta_a(1)\zeta_a(5)^3
+\frac{39248865152}{225}\zeta_a(3)\zeta_a(5)^3
+\frac{131240170112}{315}\zeta_a(5)^4
\nn\\&
+\frac{8410963}{175}\zeta_a(7)
-\frac{48485428}{15}\zeta_a(1)\zeta_a(7)
+\frac{2917402966}{105}\zeta_a(3)\zeta_a(7)
\nn\\&
-\frac{592859592}{5}\zeta_a(1)\zeta_a(3)\zeta_a(7)
-\frac{50709834308}{225}\zeta_a(3)^2\zeta_a(7)
-\frac{614764736}{15}\zeta_a(3)^3\zeta_a(7)
\nn\\&
+\frac{10528228348}{175}\zeta_a(5)\zeta_a(7)
-\frac{13704024152}{75}\zeta_a(1)\zeta_a(5)\zeta_a(7)
-\frac{636236891368}{225}\zeta_a(3)\zeta_a(5)\zeta_a(7)
\nn\\&
+\frac{1229529472}{5}\zeta_a(1)\zeta_a(3)\zeta_a(5)\zeta_a(7)
-\frac{173816402816}{225}\zeta_a(3)^2\zeta_a(5)\zeta_a(7)
\nn\\&
+\frac{486319194236}{1575}\zeta_a(5)^2\zeta_a(7)
-\frac{44855845888}{75}\zeta_a(1)\zeta_a(5)^2\zeta_a(7)
\nn\\&
-\frac{3477864507968}{1575}\zeta_a(3)\zeta_a(5)^2\zeta_a(7)
+\frac{34452188800}{63}\zeta_a(5)^3\zeta_a(7)
-\frac{4201969957}{5}\zeta_a(7)^2
\nn\\&
+\frac{33115214600}{3}\zeta_a(1)\zeta_a(7)^2
-\frac{354699152416}{75}\zeta_a(3)\zeta_a(7)^2
+\frac{19624432576}{3}\zeta_a(1)\zeta_a(3)\zeta_a(7)^2
\nn\\&
+\frac{26845490912}{45}\zeta_a(3)^2\zeta_a(7)^2
-\frac{1290516278296}{225}\zeta_a(5)\zeta_a(7)^2
\nn\\&
+13919290560\zeta_a(1)\zeta_a(5)\zeta_a(7)^2
-\frac{53491369856}{15}\zeta_a(3)\zeta_a(5)\zeta_a(7)^2
-\frac{28936277356}{45}\zeta_a(5)^2\zeta_a(7)^2
\nn\\&
-\frac{12971790427064}{225}\zeta_a(7)^3
+\frac{920228265152}{15}\zeta_a(1)\zeta_a(7)^3
+\frac{405105650152}{45}\zeta_a(3)\zeta_a(7)^3
\nn\\&
+\frac{102614104628}{45}\zeta_a(5)\zeta_a(7)^3
+\frac{10916854697}{9}\zeta_a(7)^4
-\frac{60826244}{225}\zeta_a(9)
\nn\\&
+\frac{1719513824}{75}\zeta_a(1)\zeta_a(9)
-\frac{67497730712}{225}\zeta_a(3)\zeta_a(9)
+\frac{34243973008}{25}\zeta_a(1)\zeta_a(3)\zeta_a(9)
\nn\\&
+\frac{1178861091464}{225}\zeta_a(3)^2\zeta_a(9)
+\frac{39248865152}{25}\zeta_a(3)^3\zeta_a(9)
+\frac{317821829204}{225}\zeta_a(5)\zeta_a(9)
\nn\\&
-\frac{1437536537584}{75}\zeta_a(1)\zeta_a(5)\zeta_a(9)
+\frac{753213418592}{75}\zeta_a(3)\zeta_a(5)\zeta_a(9)
\nn\\&
-\frac{235493190912}{25}\zeta_a(1)\zeta_a(3)\zeta_a(5)\zeta_a(9)
+5302586880\zeta_a(3)^2\zeta_a(5)\zeta_a(9) \nn
\end{align}
\begin{align}
&+\frac{1067731190704}{75}\zeta_a(5)^2\zeta_a(9)
-\frac{2123683755776}{75}\zeta_a(1)\zeta_a(5)^2\zeta_a(9)
\nn\\&
+\frac{79782897536}{45}\zeta_a(3)\zeta_a(5)^2\zeta_a(9)
+\frac{88186598144}{45}\zeta_a(5)^3\zeta_a(9)
+\frac{52758906296}{25}\zeta_a(7)\zeta_a(9)
\nn\\&
-\frac{691305241944}{25}\zeta_a(1)\zeta_a(7)\zeta_a(9)
-\frac{567877188992}{75}\zeta_a(3)\zeta_a(7)\zeta_a(9)
\nn\\&
-\frac{131240170112}{15}\zeta_a(1)\zeta_a(3)\zeta_a(7)\zeta_a(9)
+\frac{2402580698048}{225}\zeta_a(3)^2\zeta_a(7)\zeta_a(9)
\nn\\&
+\frac{1770759365552}{9}\zeta_a(5)\zeta_a(7)\zeta_a(9)
-\frac{982785068288}{5}\zeta_a(1)\zeta_a(5)\zeta_a(7)\zeta_a(9)
\nn\\&
-\frac{7600549533136}{225}\zeta_a(3)\zeta_a(5)\zeta_a(7)\zeta_a(9)
-\frac{410456418512}{45}\zeta_a(5)^2\zeta_a(7)\zeta_a(9)
\nn\\&
+\frac{238566772472}{225}\zeta_a(7)^2\zeta_a(9)
+\frac{27006894488}{3}\zeta_a(1)\zeta_a(7)^2\zeta_a(9)
-\frac{102614104628}{75}\zeta_a(3)\zeta_a(7)^2\zeta_a(9)
\nn\\&
-6238202684\zeta_a(5)\zeta_a(7)^2\zeta_a(9)
-\frac{27926279408}{3}\zeta_a(9)^2
+\frac{4279493160992}{25}\zeta_a(1)\zeta_a(9)^2
\nn\\&
-\frac{27279724049552}{75}\zeta_a(3)\zeta_a(9)^2
+\frac{18531431776256}{75}\zeta_a(1)\zeta_a(3)\zeta_a(9)^2
\nn\\&
+\frac{2083397615712}{25}\zeta_a(3)^2\zeta_a(9)^2
+\frac{2493810023296}{75}\zeta_a(5)\zeta_a(9)^2
-\frac{7098978088832}{75}\zeta_a(1)\zeta_a(5)\zeta_a(9)^2
\nn\\&
+\frac{703639574592}{25}\zeta_a(3)\zeta_a(5)\zeta_a(9)^2
+3564687248\zeta_a(5)^2\zeta_a(9)^2
+\frac{230934735528}{25}\zeta_a(7)\zeta_a(9)^2
\nn\\&
-\frac{820912837024}{25}\zeta_a(1)\zeta_a(7)\zeta_a(9)^2
+\frac{74858432208}{5}\zeta_a(3)\zeta_a(7)\zeta_a(9)^2
+\frac{1555643857824}{25}\zeta_a(9)^3
\nn\\&
-\frac{449150593248}{5}\zeta_a(1)\zeta_a(9)^3
+\frac{1018896758}{1575}\zeta_a(11)
-\frac{920450344}{15}\zeta_a(1)\zeta_a(11)
\nn\\&
+\frac{70243919482}{75}\zeta_a(3)\zeta_a(11)
-\frac{22512235768}{5}\zeta_a(1)\zeta_a(3)\zeta_a(11)
-\frac{1561203664196}{75}\zeta_a(3)^2\zeta_a(11)
\nn\\&
-7291056960\zeta_a(3)^3\zeta_a(11)
-\frac{3398241821656}{525}\zeta_a(5)\zeta_a(11)
+\frac{6603746363368}{75}\zeta_a(1)\zeta_a(5)\zeta_a(11)
\nn\\&
-\frac{1262141540704}{225}\zeta_a(3)\zeta_a(5)\zeta_a(11)
+43746341760\zeta_a(1)\zeta_a(3)\zeta_a(5)\zeta_a(11)
\nn\\&
-\frac{3390800471872}{225}\zeta_a(3)^2\zeta_a(5)\zeta_a(11)
-\frac{11575957519148}{45}\zeta_a(5)^2\zeta_a(11)
\nn\\&
+239349704704\zeta_a(1)\zeta_a(5)^2\zeta_a(11)
+\frac{1788524575288}{45}\zeta_a(3)\zeta_a(5)^2\zeta_a(11) \nn
\nn\\&
+\frac{322501471688}{21}\zeta_a(5)^3\zeta_a(11)
+\frac{1017951184624}{75}\zeta_a(7)\zeta_a(11)
\nn\\&
-\frac{1266334453384}{5}\zeta_a(1)\zeta_a(7)\zeta_a(11)
+\frac{125851191835072}{225}\zeta_a(3)\zeta_a(7)\zeta_a(11)
\nn\\&
-\frac{5584813454528}{15}\zeta_a(1)\zeta_a(3)\zeta_a(7)\zeta_a(11)
-\frac{5517143783536}{45}\zeta_a(3)^2\zeta_a(7)\zeta_a(11)
\nn\\&
-\frac{532196882008}{5}\zeta_a(5)\zeta_a(7)\zeta_a(11)
+\frac{848791621216}{5}\zeta_a(1)\zeta_a(5)\zeta_a(7)\zeta_a(11)
\nn\\&
-\frac{322501471688}{9}\zeta_a(3)\zeta_a(5)\zeta_a(7)\zeta_a(11)
+\frac{98028899320}{9}\zeta_a(5)^2\zeta_a(7)\zeta_a(11) \nn
\end{align}
\begin{align}
&
-\frac{1416964767526}{45}\zeta_a(7)^2\zeta_a(11)
+\frac{1128755150908}{15}\zeta_a(1)\zeta_a(7)^2\zeta_a(11)
\nn\\&
-\frac{68620229524}{3}\zeta_a(3)\zeta_a(7)^2\zeta_a(11)
+\frac{426337691032}{75}\zeta_a(9)\zeta_a(11)
\nn\\&
-\frac{2581709741336}{25}\zeta_a(1)\zeta_a(9)\zeta_a(11)
+\frac{15442054902296}{75}\zeta_a(3)\zeta_a(9)\zeta_a(11)
\nn\\&
-\frac{2121980807408}{15}\zeta_a(1)\zeta_a(3)\zeta_a(9)\zeta_a(11)
-\frac{967504415064}{25}\zeta_a(3)^2\zeta_a(9)\zeta_a(11)
\nn\\&
+65166179496\zeta_a(5)\zeta_a(9)\zeta_a(11)
-\frac{645002943376}{15}\zeta_a(1)\zeta_a(5)\zeta_a(9)\zeta_a(11)
\nn\\&
-39211559728\zeta_a(3)\zeta_a(5)\zeta_a(9)\zeta_a(11)
-\frac{950671246448}{5}\zeta_a(7)\zeta_a(9)\zeta_a(11)
\nn\\&
+274480918096\zeta_a(1)\zeta_a(7)\zeta_a(9)\zeta_a(11)
-\frac{102479366836\zeta_a(11)^2}{15}
\nn\\&
+136592436552\zeta_a(1)\zeta_a(11)^2
-358413987228\zeta_a(3)\zeta_a(11)^2
+\frac{3547516188568}{15}\zeta_a(1)\zeta_a(3)\zeta_a(11)^2
\nn\\&
+107831789252\zeta_a(3)^2\zeta_a(11)^2
+\frac{746955979352}{3}\zeta_a(5)\zeta_a(11)^2
-\frac{1078317892520}{3}\zeta_a(1)\zeta_a(5)\zeta_a(11)^2
\nn\\&
-\frac{10511644}{25}\zeta_a(13)
+\frac{1037194444}{25}\zeta_a(1)\zeta_a(13)
-\frac{116268341904}{175}\zeta_a(3)\zeta_a(13)
\nn\\&
+\frac{81238018576}{25}\zeta_a(1)\zeta_a(3)\zeta_a(13)
+\frac{711237975448}{45}\zeta_a(3)^2\zeta_a(13)
\nn\\&
+\frac{1296501688768}{225}\zeta_a(3)^3\zeta_a(13)
+\frac{2637632999144}{525}\zeta_a(5)\zeta_a(13)
-\frac{1032364965632}{15}\zeta_a(1)\zeta_a(5)\zeta_a(13)
\nn\\&
-\frac{44214157416}{25}\zeta_a(3)\zeta_a(5)\zeta_a(13)
-\frac{2593003377536}{75}\zeta_a(1)\zeta_a(3)\zeta_a(5)\zeta_a(13)
\nn\\&
+\frac{788163648272}{75}\zeta_a(3)^2\zeta_a(5)\zeta_a(13)
+\frac{75784411915784}{315}\zeta_a(5)^2\zeta_a(13)
\nn\\&
-\frac{3152654593088}{15}\zeta_a(1)\zeta_a(5)^2\zeta_a(13)
-\frac{4192519131944}{105}\zeta_a(3)\zeta_a(5)^2\zeta_a(13)
\nn\\&
-\frac{1274375691160}{63}\zeta_a(5)^3\zeta_a(13)
-\frac{1056545004944}{75}\zeta_a(7)\zeta_a(13)
+\frac{19436480109904}{75}\zeta_a(1)\zeta_a(7)\zeta_a(13)
\nn\\&
-\frac{121276155838552}{225}\zeta_a(3)\zeta_a(7)\zeta_a(13)
+\frac{5517145537904}{15}\zeta_a(1)\zeta_a(3)\zeta_a(7)\zeta_a(13)
\nn\\&
+\frac{8385038263888}{75}\zeta_a(3)^2\zeta_a(7)\zeta_a(13)
-\frac{3449523877228}{45}\zeta_a(5)\zeta_a(7)\zeta_a(13)
\nn\\&
+\frac{254875138232}{3}\zeta_a(3)\zeta_a(5)\zeta_a(7)\zeta_a(13)
+\frac{3089681550956}{15}\zeta_a(7)^2\zeta_a(13)
\nn\\&
-\frac{892062983812}{3}\zeta_a(1)\zeta_a(7)^2\zeta_a(13)
+\frac{242223957976}{25}\zeta_a(9)\zeta_a(13)
\nn\\&
-\frac{968564550096}{5}\zeta_a(1)\zeta_a(9)\zeta_a(13)
+\frac{2541481000344}{5}\zeta_a(3)\zeta_a(9)\zeta_a(13)
\nn\\&
-\frac{8385038263888}{25}\zeta_a(1)\zeta_a(3)\zeta_a(9)\zeta_a(13)
-\frac{764625414696}{5}\zeta_a(3)^2\zeta_a(9)\zeta_a(13)
\nn\\&
-\frac{1765532314832}{5}\zeta_a(5)\zeta_a(9)\zeta_a(13)
+509750276464\zeta_a(1)\zeta_a(5)\zeta_a(9)\zeta_a(13)
\nn\\&
=4.0574950525533828893641205447965377783319204691017 \cdots\times10^{-15}. \nn
\end{align}
%
\begin{align}
&\zcor{j}{j+7}=
\nn\\&
\frac{1}{12}
-\frac{49}{3}\zeta_a(1)
+\frac{3514}{9}\zeta_a(3)
-\frac{39536}{9}\zeta_a(1)\zeta_a(3)
-\frac{2535896}{45}\zeta_a(3)^2
-\frac{5575072}{15}\zeta_a(3)^3
\nn\\&
-\frac{16982}{3}\zeta_a(5)
+\frac{11053904}{45}\zeta_a(1)\zeta_a(5)
+\frac{378296}{5}\zeta_a(3)\zeta_a(5)
+\frac{11150144}{5}\zeta_a(1)\zeta_a(3)\zeta_a(5)
\nn\\&
-\frac{392354048}{135}\zeta_a(3)^2\zeta_a(5)
-15020592\zeta_a(5)^2
+\frac{3422261504}{45}\zeta_a(1)\zeta_a(5)^2
\nn\\&
+\frac{7675154528}{75}\zeta_a(3)\zeta_a(5)^2
+\frac{1522562048}{75}\zeta_a(3)^2\zeta_a(5)^2
+\frac{559650299776}{675}\zeta_a(5)^3
\nn\\&
-\frac{3045124096}{25}\zeta_a(1)\zeta_a(5)^3
+\frac{29189526016}{135}\zeta_a(3)\zeta_a(5)^3
+\frac{23245490176}{45}\zeta_a(5)^4
\nn\\&
+\frac{2224342}{45}\zeta_a(7)
-\frac{34135976}{9}\zeta_a(1)\zeta_a(7)
+\frac{171980304}{5}\zeta_a(3)\zeta_a(7)
\nn\\&
-\frac{1318776704}{9}\zeta_a(1)\zeta_a(3)\zeta_a(7)
-\frac{37730316736}{135}\zeta_a(3)^2\zeta_a(7)
-\frac{761281024}{15}\zeta_a(3)^3\zeta_a(7)
\nn\\&
+\frac{631720408}{9}\zeta_a(5)\zeta_a(7)
-\frac{9635509888}{45}\zeta_a(1)\zeta_a(5)\zeta_a(7)
-\frac{471648114368}{135}\zeta_a(3)\zeta_a(5)\zeta_a(7)
\nn\\&
+\frac{1522562048}{5}\zeta_a(1)\zeta_a(3)\zeta_a(5)\zeta_a(7)
-\frac{129267900928}{135}\zeta_a(3)^2\zeta_a(5)\zeta_a(7)
\nn\\&
+\frac{89131663936}{225}\zeta_a(5)^2\zeta_a(7)
-\frac{33359458304}{45}\zeta_a(1)\zeta_a(5)^2\zeta_a(7)
-\frac{616005489664}{225}\zeta_a(3)\zeta_a(5)^2\zeta_a(7)
\nn\\&
+\frac{2034044416}{3}\zeta_a(5)^3\zeta_a(7)
-\frac{5205948496}{5}\zeta_a(7)^2
+13664070784\zeta_a(1)\zeta_a(7)^2
\nn\\&
-\frac{791245612864}{135}\zeta_a(3)\zeta_a(7)^2
+\frac{72973815040}{9}\zeta_a(1)\zeta_a(3)\zeta_a(7)^2
+\frac{99867131392}{135}\zeta_a(3)^2\zeta_a(7)^2
\nn\\&
-\frac{953354560256}{135}\zeta_a(5)\zeta_a(7)^2
+\frac{776576324608}{45}\zeta_a(1)\zeta_a(5)\zeta_a(7)^2
-\frac{596873842432}{135}\zeta_a(3)\zeta_a(5)\zeta_a(7)^2
\nn\\&
-\frac{11958566144}{15}\zeta_a(5)^2\zeta_a(7)^2
-\frac{356964363392}{5}\zeta_a(7)^3
+\frac{684531812608}{9}\zeta_a(1)\zeta_a(7)^3
\nn\\&
+\frac{502244524544}{45}\zeta_a(3)\zeta_a(7)^3
+\frac{76330541056}{27}\zeta_a(5)\zeta_a(7)^3
+\frac{203017013248}{135}\zeta_a(7)^4
\nn\\&
-\frac{3558058}{15}\zeta_a(9)
+\frac{369424832}{15}\zeta_a(1)\zeta_a(9)
-\frac{15959058496}{45}\zeta_a(3)\zeta_a(9)
\nn\\&
+\frac{24919615168}{15}\zeta_a(1)\zeta_a(3)\zeta_a(9)
+\frac{96792887552}{15}\zeta_a(3)^2\zeta_a(9)
+\frac{29189526016}{15}\zeta_a(3)^3\zeta_a(9)
\nn\\&
+\frac{395222195008}{225}\zeta_a(5)\zeta_a(9)
-\frac{213444743168}{9}\zeta_a(1)\zeta_a(5)\zeta_a(9)
+\frac{2778439382528}{225}\zeta_a(3)\zeta_a(5)\zeta_a(9)
\nn\\&
-\frac{58379052032}{5}\zeta_a(1)\zeta_a(3)\zeta_a(5)\zeta_a(9)
+\frac{887515799552}{135}\zeta_a(3)^2\zeta_a(5)\zeta_a(9)
\nn\\&
+\frac{3954465611264}{225}\zeta_a(5)^2\zeta_a(9)
-\frac{7898847408128}{225}\zeta_a(1)\zeta_a(5)^2\zeta_a(9)
\nn\\&
+\frac{296734097408}{135}\zeta_a(3)\zeta_a(5)^2\zeta_a(9)
+\frac{328003272704}{135}\zeta_a(5)^3\zeta_a(9)
\nn\\&
+\frac{116351416496}{45}\zeta_a(7)\zeta_a(9)
-\frac{1530996228608}{45}\zeta_a(1)\zeta_a(7)\zeta_a(9)
\nn\\&
-\frac{86041013632}{9}\zeta_a(3)\zeta_a(7)\zeta_a(9)
-\frac{162718431232}{15}\zeta_a(1)\zeta_a(3)\zeta_a(7)\zeta_a(9)
\nn
\end{align}
\begin{align}
&
+\frac{357451641344}{27}\zeta_a(3)^2\zeta_a(7)\zeta_a(9)
+\frac{18274214753536}{75}\zeta_a(5)\zeta_a(7)\zeta_a(9)
\nn\\&
-\frac{3655320761344}{15}\zeta_a(1)\zeta_a(5)\zeta_a(7)\zeta_a(9)
-\frac{28269107915776}{675}\zeta_a(3)\zeta_a(5)\zeta_a(7)\zeta_a(9)
\nn\\&
-\frac{305322164224}{27}\zeta_a(5)^2\zeta_a(7)\zeta_a(9)
+\frac{19852294336}{15}\zeta_a(7)^2\zeta_a(9)
\nn\\&
+\frac{100445854208}{9}\zeta_a(1)\zeta_a(7)^2\zeta_a(9)
-\frac{76330541056}{45}\zeta_a(3)\zeta_a(7)^2\zeta_a(9)
\nn\\&
-\frac{116009721856}{15}\zeta_a(5)\zeta_a(7)^2\zeta_a(9)
-\frac{882818384896}{75}\zeta_a(9)^2
\nn\\&
+\frac{1066887023104}{5}\zeta_a(1)\zeta_a(9)^2
-\frac{101428470493952}{225}\zeta_a(3)\zeta_a(9)^2
\nn\\&
+\frac{13784809228288}{45}\zeta_a(1)\zeta_a(3)\zeta_a(9)^2
+\frac{2582952228864}{25}\zeta_a(3)^2\zeta_a(9)^2
\nn\\&
+\frac{9232723813376}{225}\zeta_a(5)\zeta_a(9)^2
-\frac{26403380928512}{225}\zeta_a(1)\zeta_a(5)\zeta_a(9)^2
\nn\\&
+\frac{174469808128}{5}\zeta_a(3)\zeta_a(5)\zeta_a(9)^2
+\frac{66291269632}{15}\zeta_a(5)^2\zeta_a(9)^2
\nn\\&
+\frac{852014044672}{75}\zeta_a(7)\zeta_a(9)^2
-\frac{610644328448}{15}\zeta_a(1)\zeta_a(7)\zeta_a(9)^2
\nn\\&
+\frac{464038887424}{25}\zeta_a(3)\zeta_a(7)\zeta_a(9)^2
+\frac{1925077604352}{25}\zeta_a(9)^3
\nn\\&
-\frac{2784233324544}{25}\zeta_a(1)\zeta_a(9)^3
+\frac{7747036}{15}\zeta_a(11)
-\frac{188223376}{3}\zeta_a(1)\zeta_a(11)
\nn\\&
+\frac{48753700336}{45}\zeta_a(3)\zeta_a(11)
-\frac{16241268736}{3}\zeta_a(1)\zeta_a(3)\zeta_a(11)
-25564928576\zeta_a(3)^2\zeta_a(11)
\nn\\&
-\frac{1220334224384}{135}\zeta_a(3)^3\zeta_a(11)
-\frac{71623304368}{9}\zeta_a(5)\zeta_a(11)
\nn\\&
+\frac{4884970955392}{45}\zeta_a(1)\zeta_a(5)\zeta_a(11)
-\frac{294936657664}{45}\zeta_a(3)\zeta_a(5)\zeta_a(11)
\nn\\&
+\frac{2440668448768}{45}\zeta_a(1)\zeta_a(3)\zeta_a(5)\zeta_a(11)
-\frac{2522348668928}{135}\zeta_a(3)^2\zeta_a(5)\zeta_a(11)
\nn\\&
-\frac{4778117440448}{15}\zeta_a(5)^2\zeta_a(11)
+\frac{2670667343872}{9}\zeta_a(1)\zeta_a(5)^2\zeta_a(11)
\nn\\&
+\frac{6652114129408}{135}\zeta_a(3)\zeta_a(5)^2\zeta_a(11)
+\frac{171354275840}{9}\zeta_a(5)^3\zeta_a(11)
\nn\\&
+\frac{773679284224}{45}\zeta_a(7)\zeta_a(11)
-\frac{2842452964736}{9}\zeta_a(1)\zeta_a(7)\zeta_a(11)
\nn\\&
+\frac{31193546502272}{45}\zeta_a(3)\zeta_a(7)\zeta_a(11)
-\frac{4154319690752}{9}\zeta_a(1)\zeta_a(3)\zeta_a(7)\zeta_a(11)
\nn\\&
-\frac{20520084530176}{135}\zeta_a(3)^2\zeta_a(7)\zeta_a(11)
-\frac{5922164213504}{45}\zeta_a(5)\zeta_a(7)\zeta_a(11)
\nn\\&
+\frac{3156941613056}{15}\zeta_a(1)\zeta_a(5)\zeta_a(7)\zeta_a(11)
-\frac{1199479930880}{27}\zeta_a(3)\zeta_a(5)\zeta_a(7)\zeta_a(11)
\nn\\&
+\frac{364601982976}{27}\zeta_a(5)^2\zeta_a(7)\zeta_a(11)
-\frac{1748582439296}{45}\zeta_a(7)^2\zeta_a(11)
\nn
\end{align}
\begin{align}
&
+\frac{839635951616}{9}\zeta_a(1)\zeta_a(7)^2\zeta_a(11)
-\frac{1276106940416}{45}\zeta_a(3)\zeta_a(7)^2\zeta_a(11)
\nn\\&
+\frac{21514347008}{3}\zeta_a(9)\zeta_a(11)
-\frac{643003794048}{5}\zeta_a(1)\zeta_a(9)\zeta_a(11)
\nn\\&
+\frac{11477049229184}{45}\zeta_a(3)\zeta_a(9)\zeta_a(11)
-\frac{7892389987328}{45}\zeta_a(1)\zeta_a(3)\zeta_a(9)\zeta_a(11)
\nn\\&
-\frac{239895986176}{5}\zeta_a(3)^2\zeta_a(9)\zeta_a(11)
+\frac{3637192835584}{45}\zeta_a(5)\zeta_a(9)\zeta_a(11)
\nn\\&
-\frac{479791972352}{9}\zeta_a(1)\zeta_a(5)\zeta_a(9)\zeta_a(11)
-\frac{729203965952}{15}\zeta_a(3)\zeta_a(5)\zeta_a(9)\zeta_a(11)
\nn\\&
-\frac{3529308941312}{15}\zeta_a(7)\zeta_a(9)\zeta_a(11)
+\frac{5104427761664}{15}\zeta_a(1)\zeta_a(7)\zeta_a(9)\zeta_a(11)
\nn\\&
-\frac{130730126912}{15}\zeta_a(11)^2
+170752598016\zeta_a(1)\zeta_a(11)^2
-\frac{20004560595712}{45}\zeta_a(3)\zeta_a(11)^2
\nn\\&
+\frac{2638855847936}{9}\zeta_a(1)\zeta_a(3)\zeta_a(11)^2
+\frac{2005310906368}{15}\zeta_a(3)^2\zeta_a(11)^2
\nn\\&
+\frac{2773028453888}{9}\zeta_a(5)\zeta_a(11)^2
-\frac{4010621812736}{9}\zeta_a(1)\zeta_a(5)\zeta_a(11)^2
-\frac{1617044}{5}\zeta_a(13)
\nn\\&
+41664480\zeta_a(1)\zeta_a(13)
-\frac{11446891456}{15}\zeta_a(3)\zeta_a(13)
+3896756864\zeta_a(1)\zeta_a(3)\zeta_a(13)
\nn\\&
+\frac{174561422464}{9}\zeta_a(3)^2\zeta_a(13)
+\frac{964503727616}{135}\zeta_a(3)^3\zeta_a(13)
+\frac{462087152384}{75}\zeta_a(5)\zeta_a(13)
\nn\\&
-\frac{1271694634496}{15}\zeta_a(1)\zeta_a(5)\zeta_a(13)
-\frac{562961649536}{225}\zeta_a(3)\zeta_a(5)\zeta_a(13)
\nn\\&
-\frac{1929007455232}{45}\zeta_a(1)\zeta_a(3)\zeta_a(5)\zeta_a(13)
+\frac{2931445783552}{225}\zeta_a(3)^2\zeta_a(5)\zeta_a(13)
\nn\\&
+\frac{13407963272704}{45}\zeta_a(5)^2\zeta_a(13)
-\frac{11725783134208}{45}\zeta_a(1)\zeta_a(5)^2\zeta_a(13)
\nn\\&
-\frac{445521117184}{9}\zeta_a(3)\zeta_a(5)^2\zeta_a(13)
-\frac{677117968384}{27}\zeta_a(5)^3\zeta_a(13)
\nn\\&
-\frac{266957923232}{15}\zeta_a(7)\zeta_a(13)
+\frac{1614470201344}{5}\zeta_a(1)\zeta_a(7)\zeta_a(13)
\nn\\&
-\frac{10017367488128}{15}\zeta_a(3)\zeta_a(7)\zeta_a(13)
+\frac{20520120484864}{45}\zeta_a(1)\zeta_a(3)\zeta_a(7)\zeta_a(13)
\nn\\&
+\frac{6237295640576}{45}\zeta_a(3)^2\zeta_a(7)\zeta_a(13)
-\frac{4286754664192}{45}\zeta_a(5)\zeta_a(7)\zeta_a(13)
\nn\\&
+\frac{4739825778688}{45}\zeta_a(3)\zeta_a(5)\zeta_a(7)\zeta_a(13)
+\frac{11470254059264}{45}\zeta_a(7)^2\zeta_a(13)
\nn\\&
-\frac{16589390225408}{45}\zeta_a(1)\zeta_a(7)^2\zeta_a(13)
+\frac{308998481792}{25}\zeta_a(9)\zeta_a(13)
\nn\\&
-\frac{1210791149568}{5}\zeta_a(1)\zeta_a(9)\zeta_a(13)
+\frac{47283506862592}{75}\zeta_a(3)\zeta_a(9)\zeta_a(13)
\nn\\&
-\frac{6237295640576}{15}\zeta_a(1)\zeta_a(3)\zeta_a(9)\zeta_a(13)
-\frac{4739825778688}{25}\zeta_a(3)^2\zeta_a(9)\zeta_a(13)
\nn\\&
-\frac{6554430891008}{15}\zeta_a(5)\zeta_a(9)\zeta_a(13)
+\frac{9479651557376}{15}\zeta_a(1)\zeta_a(5)\zeta_a(9)\zeta_a(13)
\nn\\&
=-0.022498222763372218377098609668142205119778230117126\cdots. \nn
\end{align}
%
\begin{align}
&
\zzcor{8}=
\nn\\&
\frac{1}{9}
-\frac{64}{3}\zeta_a(1)
+\frac{9728}{9}\zeta_a(3)
-\frac{58112}{3}\zeta_a(1)\zeta_a(3)
-\frac{88568768}{225}\zeta_a(3)^2
-\frac{166158848}{45}\zeta_a(3)^3
\nn\\&
-\frac{18144928}{525}\zeta_a(5)
+\frac{136763264}{75}\zeta_a(1)\zeta_a(5)
+\frac{302273408}{225}\zeta_a(3)\zeta_a(5)
+\frac{332317696}{15}\zeta_a(1)\zeta_a(3)\zeta_a(5)
\nn\\&
-\frac{6689508352}{225}\zeta_a(3)^2\zeta_a(5)
-\frac{245847510208}{1575}\zeta_a(5)^2
+\frac{4090017792}{5}\zeta_a(1)\zeta_a(5)^2
\nn\\&
+\frac{1279047013888}{1125}\zeta_a(3)\zeta_a(5)^2
+\frac{771863332864}{3375}\zeta_a(3)^2\zeta_a(5)^2
+\frac{73210491209728}{7875}\zeta_a(5)^3
\nn\\&
-\frac{1543726665728}{1125}\zeta_a(1)\zeta_a(5)^3
+\frac{1642937741312}{675}\zeta_a(3)\zeta_a(5)^3
+\frac{27470069819392}{4725}\zeta_a(5)^4
\nn\\&
+\frac{96713248}{175}\zeta_a(7)
-\frac{200023936}{5}\zeta_a(1)\zeta_a(7)
+\frac{566287640704}{1575}\zeta_a(3)\zeta_a(7)
\nn\\&
-\frac{23985625088}{15}\zeta_a(1)\zeta_a(3)\zeta_a(7)
-\frac{139076844544}{45}\zeta_a(3)^2\zeta_a(7)
-\frac{385931666432}{675}\zeta_a(3)^3\zeta_a(7)
\nn\\&
+\frac{1324465806976}{1575}\zeta_a(5)\zeta_a(7)
-\frac{64908306432}{25}\zeta_a(1)\zeta_a(5)\zeta_a(7)
-\frac{61857539616256}{1575}\zeta_a(3)\zeta_a(5)\zeta_a(7)
\nn\\&
+\frac{771863332864}{225}\zeta_a(1)\zeta_a(3)\zeta_a(5)\zeta_a(7)
-\frac{7275867140096}{675}\zeta_a(3)^2\zeta_a(5)\zeta_a(7)
\nn\\&
+\frac{3681144976896}{875}\zeta_a(5)^2\zeta_a(7)
-\frac{1877643132928}{225}\zeta_a(1)\zeta_a(5)^2\zeta_a(7)
\nn\\&
-\frac{727956850213888}{23625}\zeta_a(3)\zeta_a(5)^2\zeta_a(7)
+\frac{7211172859904}{945}\zeta_a(5)^3\zeta_a(7)
\nn\\&
-\frac{2604590283808}{225}\zeta_a(7)^2
+\frac{2298682281472}{15}\zeta_a(1)\zeta_a(7)^2
-\frac{14807437376512}{225}\zeta_a(3)\zeta_a(7)^2
\nn\\&
+\frac{821468870656}{9}\zeta_a(1)\zeta_a(3)\zeta_a(7)^2
+\frac{1873104521728}{225}\zeta_a(3)^2\zeta_a(7)^2
-\frac{18056136893696}{225}\zeta_a(5)\zeta_a(7)^2
\nn\\&
+\frac{43701512508416}{225}\zeta_a(1)\zeta_a(5)\zeta_a(7)^2
-\frac{11196205586432}{225}\zeta_a(3)\zeta_a(5)\zeta_a(7)^2
\nn\\&
-\frac{224319044096}{25}\zeta_a(5)^2\zeta_a(7)^2
-\frac{60364188889088}{75}\zeta_a(7)^3
+\frac{38522095904768}{45}\zeta_a(1)\zeta_a(7)^3
\nn\\&
+\frac{28263843908608}{225}\zeta_a(3)\zeta_a(7)^3
+\frac{4295556306944}{135}\zeta_a(5)\zeta_a(7)^3
+\frac{11424868964096}{675}\zeta_a(7)^4
\nn\\&
-\frac{892645952}{225}\zeta_a(9)
+\frac{24262339712}{75}\zeta_a(1)\zeta_a(9)
-\frac{936479578496}{225}\zeta_a(3)\zeta_a(9)
\nn\\&
+\frac{1430082970624}{75}\zeta_a(1)\zeta_a(3)\zeta_a(9)
+\frac{16406798258176}{225}\zeta_a(3)^2\zeta_a(9)
+\frac{1642937741312}{75}\zeta_a(3)^3\zeta_a(9)
\nn\\&
+\frac{21838484383936}{1125}\zeta_a(5)\zeta_a(9)
-\frac{19962425624576}{75}\zeta_a(1)\zeta_a(5)\zeta_a(9)
\nn\\&
+\frac{157984107122176}{1125}\zeta_a(3)\zeta_a(5)\zeta_a(9)
-\frac{3285875482624}{25}\zeta_a(1)\zeta_a(3)\zeta_a(5)\zeta_a(9)
\nn\\&
+\frac{49944585723904}{675}\zeta_a(3)^2\zeta_a(5)\zeta_a(9)
+\frac{223789604476928}{1125}\zeta_a(5)^2\zeta_a(9)
\nn\\&
-\frac{444505717600256}{1125}\zeta_a(1)\zeta_a(5)^2\zeta_a(9)
+\frac{16699023499264}{675}\zeta_a(3)\zeta_a(5)^2\zeta_a(9)
\nn
\end{align}
\begin{align}
&
+\frac{18458180190208}{675}\zeta_a(5)^3\zeta_a(9)
+\frac{2215265874304}{75}\zeta_a(7)\zeta_a(9)
-\frac{9672582147072}{25}\zeta_a(1)\zeta_a(7)\zeta_a(9)
\nn\\&
-\frac{23537685095936}{225}\zeta_a(3)\zeta_a(7)\zeta_a(9)
-\frac{27470069819392}{225}\zeta_a(1)\zeta_a(3)\zeta_a(7)\zeta_a(9)
\nn\\&
+\frac{100576296452096}{675}\zeta_a(3)^2\zeta_a(7)\zeta_a(9)
+\frac{1029998804728832}{375}\zeta_a(5)\zeta_a(7)\zeta_a(9)
\nn\\&
-\frac{205703916593152}{75}\zeta_a(1)\zeta_a(5)\zeta_a(7)\zeta_a(9)
-\frac{1590848686794752}{3375}\zeta_a(3)\zeta_a(5)\zeta_a(7)\zeta_a(9)
\nn\\&
-\frac{17182225227776}{135}\zeta_a(5)^2\zeta_a(7)\zeta_a(9)
+\frac{3318403631872}{225}\zeta_a(7)^2\zeta_a(9)
\nn\\&
+\frac{5652697652224}{45}\zeta_a(1)\zeta_a(7)^2\zeta_a(9)
-\frac{4295556306944}{225}\zeta_a(3)\zeta_a(7)^2\zeta_a(9)
\nn\\&
-\frac{6528496550912}{75}\zeta_a(5)\zeta_a(7)^2\zeta_a(9)
-\frac{16063303796224}{125}\zeta_a(9)^2
+\frac{178602817662976}{75}\zeta_a(1)\zeta_a(9)^2
\nn\\&
-\frac{5710190085059584}{1125}\zeta_a(3)\zeta_a(9)^2
+\frac{775749425217536}{225}\zeta_a(1)\zeta_a(3)\zeta_a(9)^2
\nn\\&
+\frac{145356454268928}{125}\zeta_a(3)^2\zeta_a(9)^2
+\frac{523156377628672}{1125}\zeta_a(5)\zeta_a(9)^2
\nn\\&
-\frac{1485862928564224}{1125}\zeta_a(1)\zeta_a(5)\zeta_a(9)^2
+\frac{9818414415872}{25}\zeta_a(3)\zeta_a(5)\zeta_a(9)^2
\nn\\&
+\frac{3730569457664}{75}\zeta_a(5)^2\zeta_a(9)^2
+\frac{48528226740224}{375}\zeta_a(7)\zeta_a(9)^2
-\frac{34364450455552}{75}\zeta_a(1)\zeta_a(7)\zeta_a(9)^2
\nn\\&
+\frac{26113986203648}{125}\zeta_a(3)\zeta_a(7)\zeta_a(9)^2
+\frac{108635330463744}{125}\zeta_a(9)^3
-\frac{156683917221888}{125}\zeta_a(1)\zeta_a(9)^3
\nn\\&
+\frac{17053628768\zeta_a(11)}{1575}
-\frac{13883064448}{15}\zeta_a(1)\zeta_a(11)
+\frac{1006601134528}{75}\zeta_a(3)\zeta_a(11)
\nn\\&
-\frac{953720255488}{15}\zeta_a(1)\zeta_a(3)\zeta_a(11)
-\frac{65579160207616}{225}\zeta_a(3)^2\zeta_a(11)
\nn\\&
-\frac{68673805370368}{675}\zeta_a(3)^3\zeta_a(11)
-\frac{3164749481344}{35}\zeta_a(5)\zeta_a(11)
+\frac{92292621735424}{75}\zeta_a(1)\zeta_a(5)\zeta_a(11)
\nn\\&
-\frac{2011639173632}{25}\zeta_a(3)\zeta_a(5)\zeta_a(11)
+\frac{137347610740736}{225}\zeta_a(1)\zeta_a(3)\zeta_a(5)\zeta_a(11)
\nn\\&
-\frac{141944312147968}{675}\zeta_a(3)^2\zeta_a(5)\zeta_a(11)
-\frac{5656297178811136}{1575}\zeta_a(5)^2\zeta_a(11)
\nn\\&
+\frac{16699184758784}{5}\zeta_a(1)\zeta_a(5)^2\zeta_a(11)
+\frac{374350251478016}{675}\zeta_a(3)\zeta_a(5)^2\zeta_a(11)
\nn\\&
+\frac{13500319821824}{63}\zeta_a(5)^3\zeta_a(11)
+\frac{4679826787328}{25}\zeta_a(7)\zeta_a(11)
-\frac{52838465354752}{15}\zeta_a(1)\zeta_a(7)\zeta_a(11)
\nn\\&
+\frac{1756254622363648}{225}\zeta_a(3)\zeta_a(7)\zeta_a(11)
-\frac{233787344924672}{45}\zeta_a(1)\zeta_a(3)\zeta_a(7)\zeta_a(11)
\nn\\&
-\frac{1154775437268992}{675}\zeta_a(3)^2\zeta_a(7)\zeta_a(11)
-\frac{111542847894016}{75}\zeta_a(5)\zeta_a(7)\zeta_a(11)
\nn\\&
+\frac{177657888550912}{75}\zeta_a(1)\zeta_a(5)\zeta_a(7)\zeta_a(11)
-\frac{13500319821824}{27}\zeta_a(3)\zeta_a(5)\zeta_a(7)\zeta_a(11)
\nn\\&
+\frac{20518132017152}{135}\zeta_a(5)^2\zeta_a(7)\zeta_a(11)
-\frac{99086064315392}{225}\zeta_a(7)^2\zeta_a(11)
\nn
\end{align}
\begin{align}
&
+\frac{47251119376384}{45}\zeta_a(1)\zeta_a(7)^2\zeta_a(11)
-\frac{71813462060032}{225}\zeta_a(3)\zeta_a(7)^2\zeta_a(11)
\nn\\&
+\frac{1965929854592}{25}\zeta_a(9)\zeta_a(11)
-\frac{107837119706624}{75}\zeta_a(1)\zeta_a(9)\zeta_a(11)
\nn\\&
+\frac{646701599847424}{225}\zeta_a(3)\zeta_a(9)\zeta_a(11)
-\frac{444145559689216}{225}\zeta_a(1)\zeta_a(3)\zeta_a(9)\zeta_a(11)
\nn\\&
-\frac{13500319821824}{25}\zeta_a(3)^2\zeta_a(9)\zeta_a(11)
+\frac{204550905540608}{225}\zeta_a(5)\zeta_a(9)\zeta_a(11)
\nn\\&
-\frac{27000639643648}{45}\zeta_a(1)\zeta_a(5)\zeta_a(9)\zeta_a(11)
-\frac{41036264034304}{75}\zeta_a(3)\zeta_a(5)\zeta_a(9)\zeta_a(11)
\nn\\&
-\frac{199164772516864}{75}\zeta_a(7)\zeta_a(9)\zeta_a(11)
+\frac{287253848240128}{75}\zeta_a(1)\zeta_a(7)\zeta_a(9)\zeta_a(11)
\nn\\&
-\frac{2349684379328}{25}\zeta_a(11)^2
+\frac{28472179501568}{15}\zeta_a(1)\zeta_a(11)^2
-\frac{1125029980473344}{225}\zeta_a(3)\zeta_a(11)^2
\nn\\&
+\frac{148503518040064}{45}\zeta_a(1)\zeta_a(3)\zeta_a(11)^2
+\frac{112849726094336}{75}\zeta_a(3)^2\zeta_a(11)^2
\nn\\&
+\frac{156486606977536}{45}\zeta_a(5)\zeta_a(11)^2
-\frac{225699452188672}{45}\zeta_a(1)\zeta_a(5)\zeta_a(11)^2
-\frac{184488304}{25}\zeta_a(13)
\nn\\&
+\frac{16006015168}{25}\zeta_a(1)\zeta_a(13)
-\frac{5050137517568}{525}\zeta_a(3)\zeta_a(13)
+\frac{3456789546496}{75}\zeta_a(1)\zeta_a(3)\zeta_a(13)
\nn\\&
+\frac{9975441242624}{45}\zeta_a(3)^2\zeta_a(13)
+\frac{54274656477184}{675}\zeta_a(3)^3\zeta_a(13)
+\frac{185004099627776}{2625}\zeta_a(5)\zeta_a(13)
\nn\\&
-\frac{24086751383552}{25}\zeta_a(1)\zeta_a(5)\zeta_a(13)
-\frac{60191060520448}{2625}\zeta_a(3)\zeta_a(5)\zeta_a(13)
\nn\\&
-\frac{108549312954368}{225}\zeta_a(1)\zeta_a(3)\zeta_a(5)\zeta_a(13)
+\frac{164968039368704}{1125}\zeta_a(3)^2\zeta_a(5)\zeta_a(13)
\nn\\&
+\frac{1763405155246592}{525}\zeta_a(5)^2\zeta_a(13)
-\frac{659872157474816}{225}\zeta_a(1)\zeta_a(5)^2\zeta_a(13)
\nn\\&
-\frac{175504157683712}{315}\zeta_a(3)\zeta_a(5)^2\zeta_a(13)
-\frac{38105102317568}{135}\zeta_a(5)^3\zeta_a(13)
\nn\\&
-\frac{14605660125056}{75}\zeta_a(7)\zeta_a(13)
+\frac{90182376920064}{25}\zeta_a(1)\zeta_a(7)\zeta_a(13)
\nn\\&
-\frac{338559015528448}{45}\zeta_a(3)\zeta_a(7)\zeta_a(13)
+\frac{1154776275580928}{225}\zeta_a(1)\zeta_a(3)\zeta_a(7)\zeta_a(13)
\nn\\&
+\frac{351008315367424}{225}\zeta_a(3)^2\zeta_a(7)\zeta_a(13)
-\frac{80128911070208}{75}\zeta_a(5)\zeta_a(7)\zeta_a(13)
\nn\\&
+\frac{266735716222976}{225}\zeta_a(3)\zeta_a(5)\zeta_a(7)\zeta_a(13)
+\frac{647285510679808}{225}\zeta_a(7)^2\zeta_a(13)
\nn\\&
-\frac{933575006780416}{225}\zeta_a(1)\zeta_a(7)^2\zeta_a(13)
+\frac{16661398326144}{125}\zeta_a(9)\zeta_a(13)
\nn\\&
-\frac{67297878821888}{25}\zeta_a(1)\zeta_a(9)\zeta_a(13)
+\frac{2659161772027904}{375}\zeta_a(3)\zeta_a(9)\zeta_a(13)
\nn\\&
-\frac{351008315367424}{75}\zeta_a(1)\zeta_a(3)\zeta_a(9)\zeta_a(13)
-\frac{266735716222976}{125}\zeta_a(3)^2\zeta_a(9)\zeta_a(13)
\nn\\&
-\frac{369877434674176}{75}\zeta_a(5)\zeta_a(9)\zeta_a(13)
+\frac{533471432445952}{75}\zeta_a(1)\zeta_a(5)\zeta_a(9)\zeta_a(13)
\nn\\&
=0.40724241475962085558975358793634670774086664004174\cdots. \nn
\end{align}

\end{appendix}


\end{document}